\documentclass[aps,prb,twocolumn,superscriptaddress,showpacs,10pt]{revtex4}

\usepackage{color}
\usepackage{amsmath}
\usepackage{verbatim}
\usepackage{hyperref}

\usepackage{graphicx}
\usepackage{overpic}

\def\Tiny{ \font\Tinyfont = cmr10 at 5.5pt \relax  \Tinyfont}

\usepackage[caption=false,lofdepth,lotdepth]{subfig}
\usepackage{dcolumn}
\usepackage{bm}

\newcommand{\ket}[1]{|#1\rangle}

\newcommand{\vect}[1]{\boldsymbol{#1}}
\newcommand{\da}{\downarrow}
\newcommand{\ua}{\uparrow}
\newcommand{\dagg}[1]{#1^{\dagger}}

\makeatletter\def\CT{\def\@captype{figure}}\makeatother
\graphicspath{{./Figures/}}

\begin{document}
\title{Theory of magnetism and metal-insulator transition in layered perovskite iridates}

\author{Jean-Michel Carter}
\affiliation{Department of Physics, University of Toronto, Toronto, Ontario M5S 1A7 Canada} 

\author{Vijay Shankar V.}
\affiliation{Department of Physics, University of Toronto, Toronto, Ontario M5S 1A7 Canada} 

\author{Hae-Young Kee}
\email{hykee@physics.utoronto.ca}
\affiliation{Department of Physics, University of Toronto, Toronto, Ontario M5S 1A7 Canada}
\affiliation{Canadian Institute for Advanced Research, Toronto, Ontario  Canada}

\begin{abstract} 
We investigate the metal-insulator transition in the layered Ruddelsden Popper series of strontium iridates Sr$_{n+1}$Ir$_n$O$_{3n+1}$. Tight-binding models of $t_{2g}$ orbitals for $n=1$, $2$, and $\infty$ are constructed, and changes in band dispersion due to dimensionality and spin-orbit coupling are presented. Identifying the states near the Fermi level to be predominantly J$_{\text{eff}}=1/2$, we use an effective Hubbard model to study the effect of correlations. Transitions from a metallic state to various magnetically ordered states at different critical interactions are obtained. A canted antiferromagnetic insulator is found for Sr$_{2}$IrO$_{4}$, a c-axis collinear antiferromagnetic insulator for Sr$_{3}$Ir$_2$O$_{7}$, and non-coplanar canted antiferromagnetic insulator via magnetic metal for SrIrO$_3$. We derive the strong-coupling spin-model and compare the magnetic ordering patterns obtained in the weak and strong coupling limits. We find that they are identical, indicating that magnetic ordering is not sufficient to justify Mott physics in this series of iridates. 

\end{abstract}


\maketitle

\section{Introduction} 
\label{sec:introduction}
Spin-orbit coupling (SOC) is often either ignored or treated perturbatively in correlated electronic materials where 3d or 4d orbitals are active.
However, in materials with heavier atoms, SOC plays an important role in determining the
ground state and often helps stabilize novel phases such as topological insulators \cite{Hasan_Kane_Review,Qi_Zhang_review} and spin liquids\cite{balents2010spin}. In particular, the interplay between SOC and electronic interactions needs to be understood when these energy scales are comparable. Iridium oxides (iridates) with 5d-orbitals offer a playground to investigate their combined effects.\cite{Lawler:2008oq, Chen-2008-Spin-orbit, Balents,Yang:2010uq,  Wan:2011nx, Jiang:2011cr,Xiao-2011-Interface,Will:PRB}

In the Ruddlesden Popper series of Iridium oxides, Sr$_{n+1}$Ir$_n$O$_{3n+1}$, octahedral crystal field due to oxygen atoms separates the $e_g$ and t$_{2g}$ states, leaving five electrons on the t$_{2g}$ levels. When SOC is large, these t$_{2g}$ states are split into J$_{\text{eff}}=1/2$ and J$_{\text{eff}} = 3/2$, and the system can be viewed as half-filled J$_{\text{eff}}=1/2$. Since the bandwidth of J$_{\text{eff}}=1/2$, $W_j$, is narrower than the original bandwidth $W_{t_{2g}}$ of all the t$_{2g}$ states without SOC, the Hubbard interaction $U$ on the half-filled J$_{\text{eff}}=1/2$ state is effectively enhanced, as the ratio $U/W_{j}$ is larger than $U/W_{t_{2g}}$. Thus
the effect of Hubbard interaction in iridates is amplified due to strong SOC via narrowing of bandwidth, leading to an insulating state in some layered perovskite\cite{Crawford:1994ys,Cao:1998uq,Cao:2002uq,Nagai:2007vn,Kim:2008ve,Moon:2008ly,Jackeli:2009qf,Kim:2009bh} and pyrochlore\cite{Balents,Yang:2010uq} iridates. This state, dubbed spin-orbit Mott insulator, was first reported in single layered perovskite Sr$_2$IrO$_4$ (Sr-214)\cite{Kim:2008ve,Moon:2008ly,Kim:2009bh}.

This narrowing of bandwidth idea was applied to explain the insulator-metal transition in Sr$_{n+1}$Ir$_n$O$_{3n+1}$ series as a function of $n$. While single layer Sr$_2$IrO$_4$ (Sr-214) with $n=1$ and the bilayer Sr$_3$Ir$_2$O$_7$ (Sr-327) with $n=2$ are insulators, three-dimensional SrIrO$_3$ (Sr-113) with $n=\infty$ is metallic\cite{Moon:2008ly}.
Since J$_{\text{eff}}=1/2$ state made of an equal mixture of d$_{xy}$, d$_{xz}$ and d$_{yz}$ orbitals is almost isotropic, Sr-113 has a larger bandwidth than the layered compounds Sr-214 and Sr-327. $U/W_j$ is smaller in Sr-113,  which then leads to a metallic state in Sr-113.
Sr-327, via analogous reasoning, is closer to a metal-insulator transition with a smaller charge gap than Sr-214.
This mechanism was referred to as dimensionality controlled metal-insulator transition (MIT)\cite{Moon:2008ly}.

The above proposal, however, is only valid when SOC is large enough to split J$_{\text{eff}}=1/2$ and $3/2$ bands. Once these two bands are well separated, $W_j$ increases as $n$ increases, and one can compare it with the Hubbard interaction $U$. In the opposite limit, without SOC, the original t$_{2g}$ bandwidth $W_{t_{2g}}$ is not very sensitive to the dimensionality. Therefore, the dimensionality controlled MIT scenario strongly depends on the strength of SOC. With this in mind, it is important to notice recent experimental and theoretical developments. Multiple angle resolved photoemission spectroscopy (ARPES) measurements \cite{ARPES:dessau-arxiv,ARPES:wojek-jpcm, Kim:2008ve}, ab-initio calculations \cite{Watanabe:2010cr,Arita:PRL}, and a tight-binding theory of Sr-214 and Sr-327\cite{Carter:327} report that these two bands are not well separated, and the mixing between J$_{\text{eff}}=1/2$ and $3/2$ is not negligible on the occupied bands. 

When SOC is intermediate, close to the values deduced from ab-initio calculations and ARPES, $W_j$ is not well-defined, and the occupied bands are a mixture of J$_{\text{eff}}=1/2$ and J$_{\text{eff}}=3/2$.
As the dimensionality $n$ is increased, 
the mixing between $1/2$ and $3/2$ among the occupied bands increases. The actual bandwidth $W$ in the case of intermediate SOC is larger than $W_{t_{2g}}$ by the effect of spin-orbit coupling.   
The estimated Hubbard $U$ is smaller than this bandwidth $W$ and a strong coupling Mott insulator scheme is questionable. It was pointed out by Arita et. al \cite{Arita:PRL} that Sr-214 is a Slater rather than a Mott insulator. Given that the magnetic structure does not break any further translational symmetry unlike a typical Slater insulator, the Slater behaviour is not immediately apparent.

In this paper, we study the MIT as a function of the number of layers in a unit cell within the Slater picture. If the Slater picture is appropriate, this approach should reveal a MIT in the layered iridates. To proceed, the states near the Fermi level rather than all the occupied states need to be identified. While the mixing between J$_{\text{eff}}=1/2$ and J$_{\text{eff}}=3/2$ is large for the occupied states, the states near the Fermi level are mainly J$_{\text{eff}}=1/2$ for the case of intermediate SOC. Thus, we carry out a  mean field study of a Hubbard model for the J$_{\text{eff}}=1/2$ bands near E$_F$. The density of states (DOS) near the Fermi level is inherited from the point group symmetry of the crystal structure, and not simply determined by the dimensionality of crystal structure. Our results imply that the combination of SOC and the crystal structure plays an important role in determining the different magnetic ordering patterns for this series of iridates. In turn, we suggest that the pattern of magnetic ordering and the size of the magnetic moment in the insulating phase can be sensitive to changes in crystal structure. While the magnetic ordering itself is a consequence of electronic correlations and should not be sensitive to the details, this does not guarantee a Mott insulator as will be discussed below.

The paper is organised as follows. In Sec. \ref{sec:t2g}, we introduce a tight-binding model with SOC using $t_{2g}$ orbitals for the different layered iridates. From this model, we identify the states near the Fermi level as being comprised of mostly J$_{\text{eff}} =\frac{1}{2}$ orbitals. Then, in Sec. \ref{sec:Model} we discuss an effective tight-binding model using J$_{\text{eff}}=\frac{1}{2}$ orbitals that reproduces the band structure near the Fermi level. We then study the effect of interactions at a mean-field level using an effective Hubbard model for these orbitals.
The magnetic states obtained using a self-consistent method are then explained in Sec. \ref{sec:magnetism-mean-field-theory}. A collinear antiferromagnetic (AF) state with moments oriented along the crystal c-axis is found in Sr-327, while a coplanar canted AF is realized in Sr-214. The full unit-cell structure in Sr-214 breaks spin-rotation symmetry and favours the canted AF state, while the bilayer structure determines the collinear AF state in Sr-327. Sr-113, although close to a magnetic transition, remains metallic because interactions in the real material is smaller than the critical interaction strength required for this magnetic transition. 
The results found at the mean-field level are compared with spin-models derived in the strong coupling limit in Sec. \ref{sec:magnetism_spin_model}. Finally, we discuss implications of our results and conclude in Sec. \ref{sec:discussion_and_summary}.
\begin{figure}[t]
\includegraphics[height=3in,width=3.35in,angle=0]{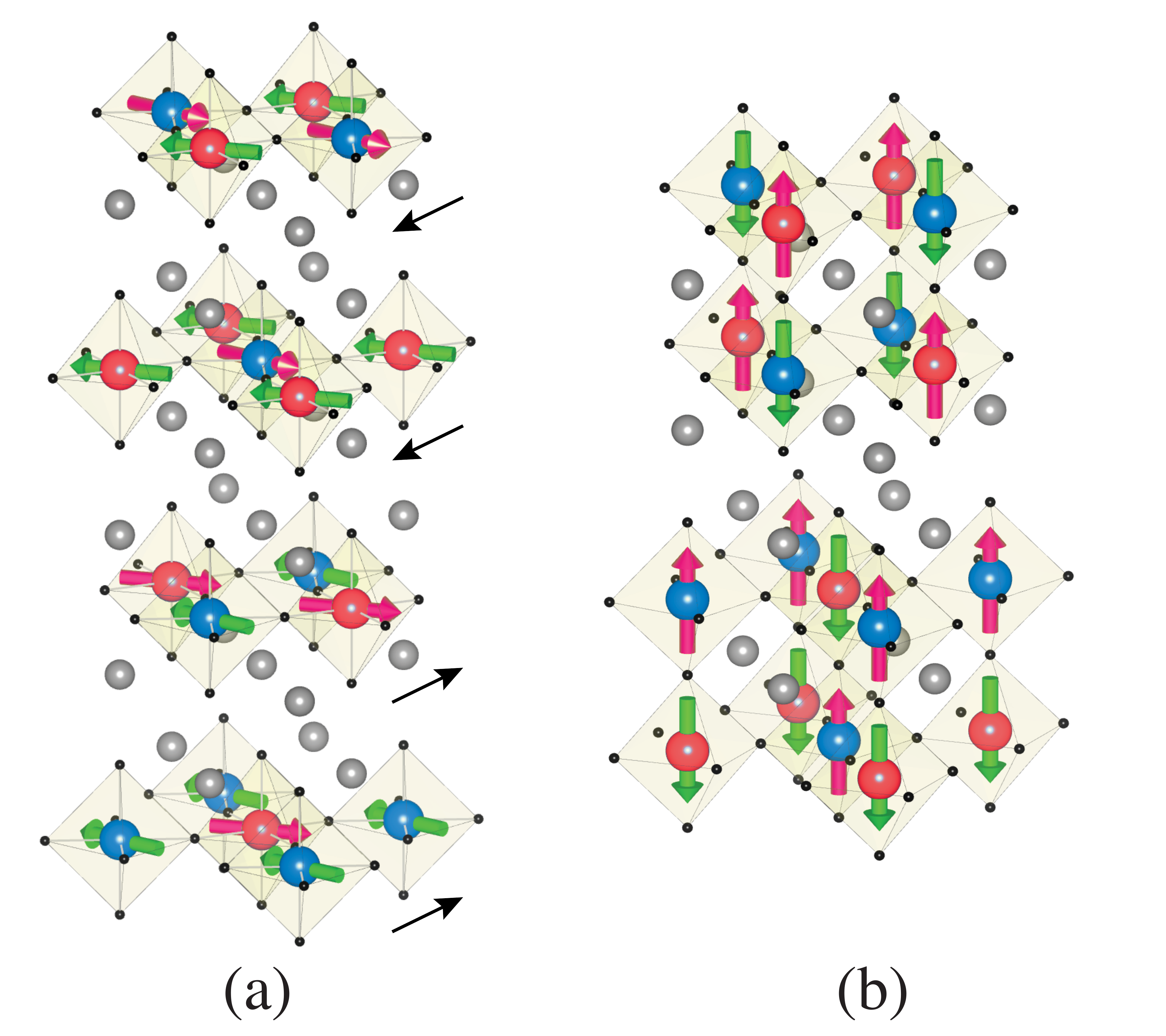}
\caption{(Color online) Theoretical  magnetic ordering in Sr-214 (a) and Sr-327 (b) in the insulating phase. Each layer has two different Iridium atoms because of the staggered rotation of the octahedra which we label blue and red. (a) The coplanar canted AF state is shown with the spin configuration for each layer. The black arrow represents the direction of the net ferromagnetic moment within each plane, showing the up-up-down-down structure. (b) The favored collinear AF state with moments pointing along the c-axis is shown; the second bilayer has its spins flipped from the first set.}
\label{Fig:Mag_ordering}
\end{figure}
\section{Tight binding model with $\text{t}_{2g}$ orbitals}
\label{sec:t2g}
In the layered Ruddelsden-Popper series of iridates, the oxidation state of Iridium is Ir$^{4+}$ with an electronic configuration of $\text{[Xe]5d}^{5}$. The octahedral crystal field from the oxygen atoms further splits the $5d$ levels into triply degenerate $t_{2g}$ and doubly degenerate $e_g$ levels with $t_{2g}$ lying lower in energy. There are then five $d$ electrons in the $t_{2g}$ levels. Iridium, being a heavy atom, has significant spin-orbit coupling. Within the low-lying $t_{2g}$ subspace, the orbital angular momentum behaves like an effective angular momentum $l=1$ with a negative spin-orbit coupling constant $\lambda$.\cite{Fazekas} With this picture of the atomic levels in Iridium, tight-binding models incorporating SOC within the t$_{2g}$ orbitals $d_{yz}$, $d_{xz}$, and $d_{xy}$ is discussed for the different layered iridates. 
\label{sec:214-t2g}
\begin{figure*}
  \begingroup
\label{fig:214-band}
\subfloat[][\label{first}$\lambda= 0t$]{     
\includegraphics[scale=0.21,grid]{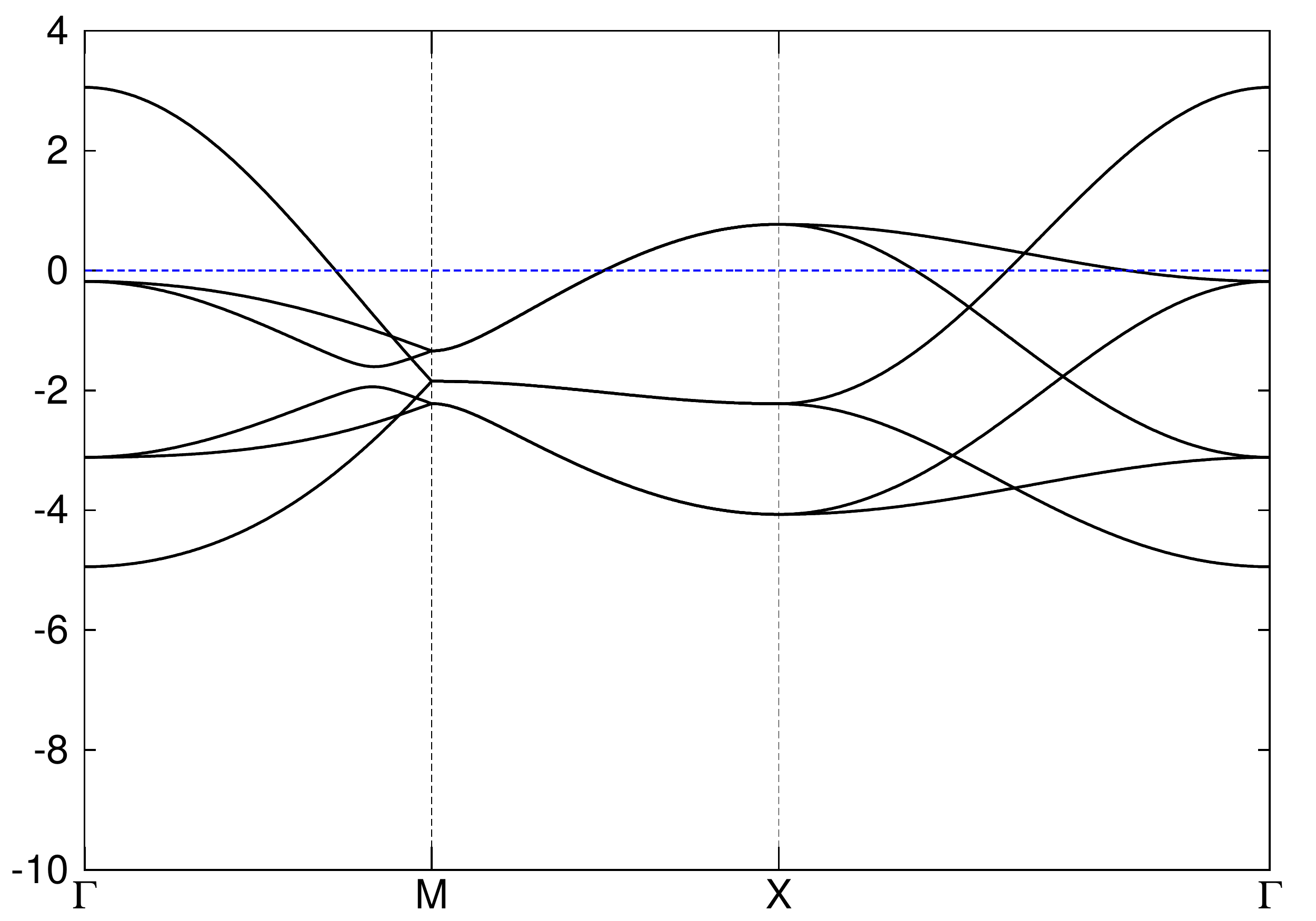}
} \quad
\subfloat[][\label{fig:214-bs-second}$\lambda=2t$]{     
 \includegraphics[scale=0.21]{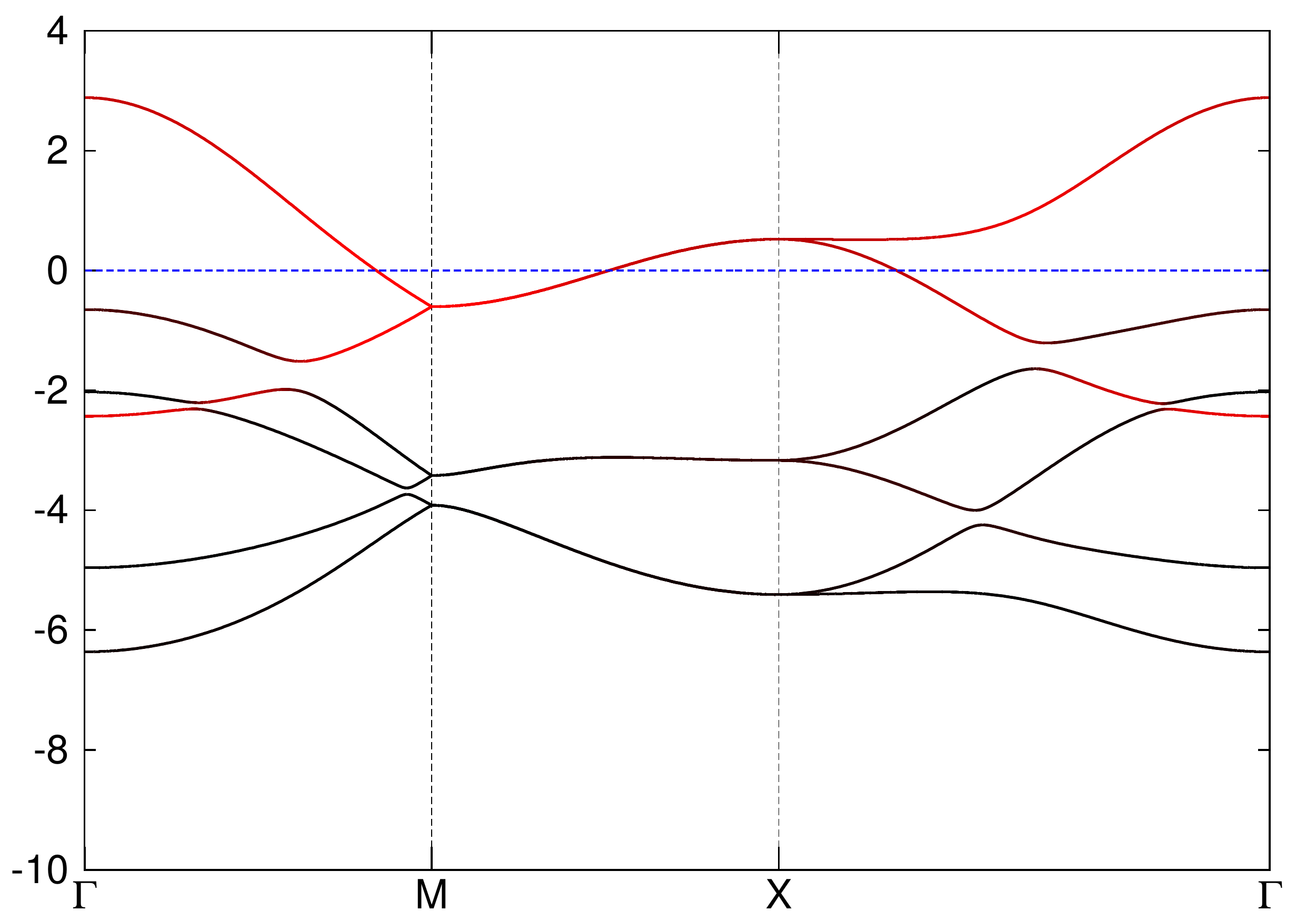}} \quad
\subfloat[][\label{third}$\lambda=4t$]{     
  \includegraphics[scale=0.21]{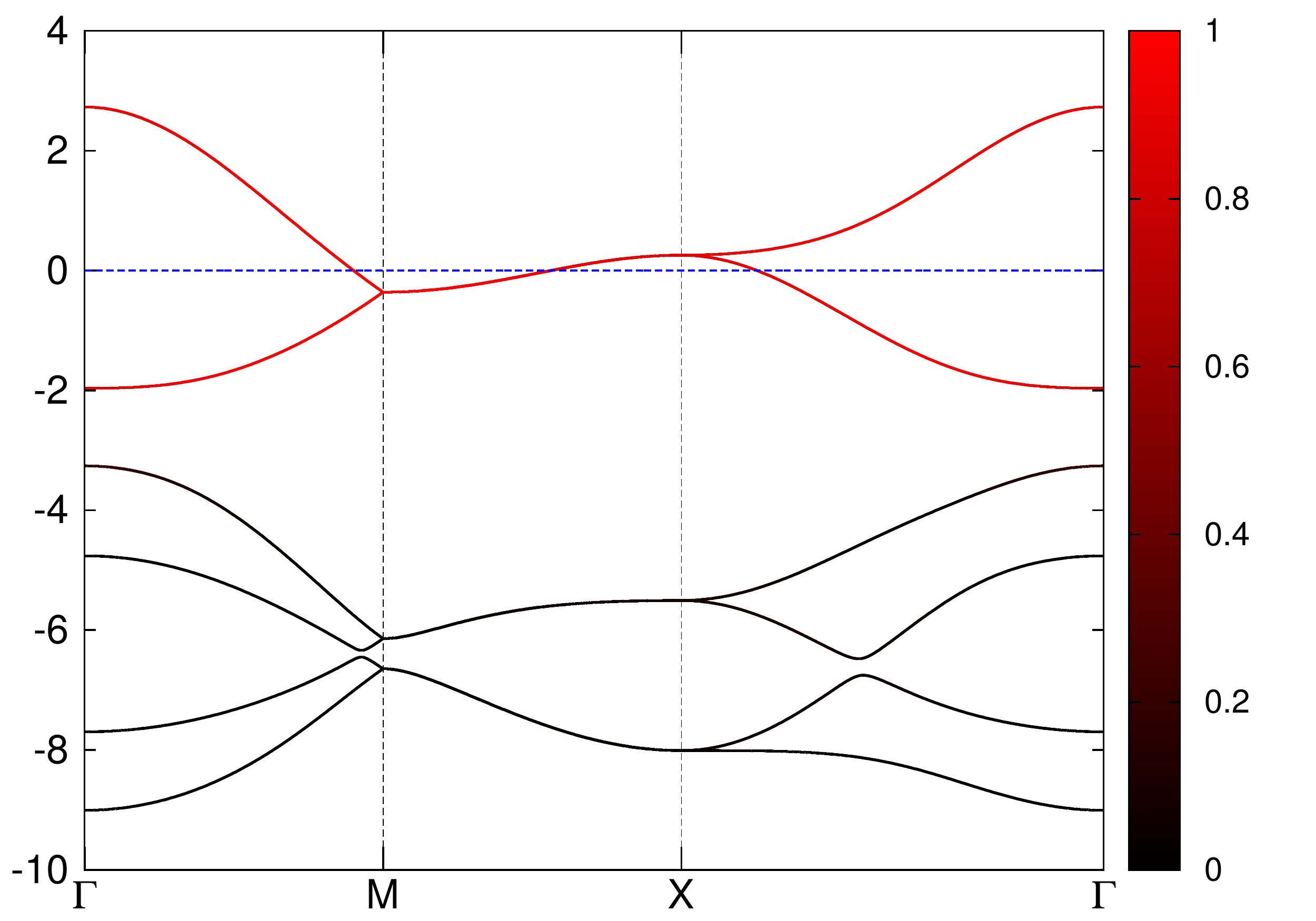}} 

\endgroup

\begingroup
\label{fig:327-band}
\subfloat[][{\label{fig:327-1}}$\lambda= 0t$]{     
 \includegraphics[scale=0.43]{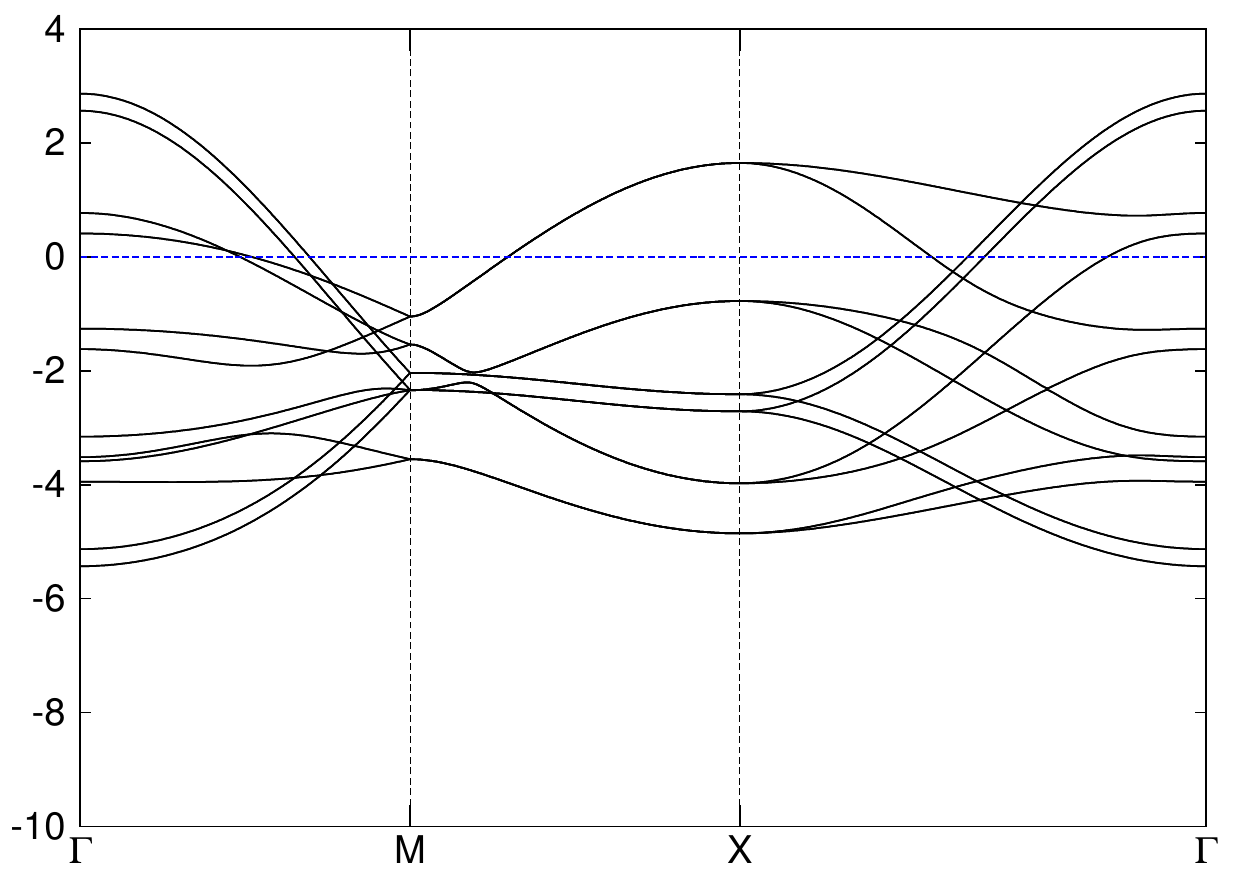}} \quad
\subfloat[][\label{fig:327-bs-second}$\lambda=2t$]{     
 \includegraphics[scale=0.43]{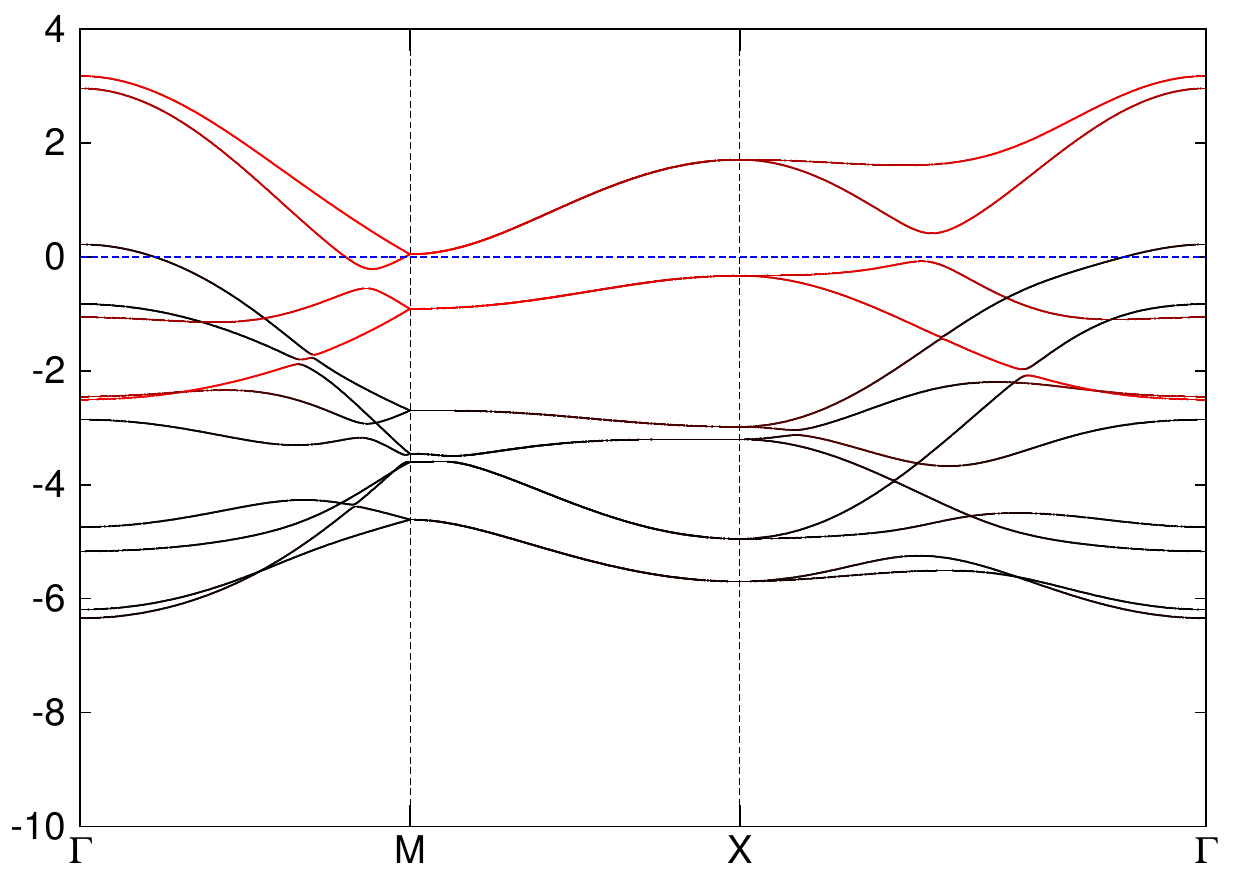}}\quad
\subfloat[][{\label{fig:327-3}}$\lambda=4t$]{     
  \includegraphics[scale=0.43]{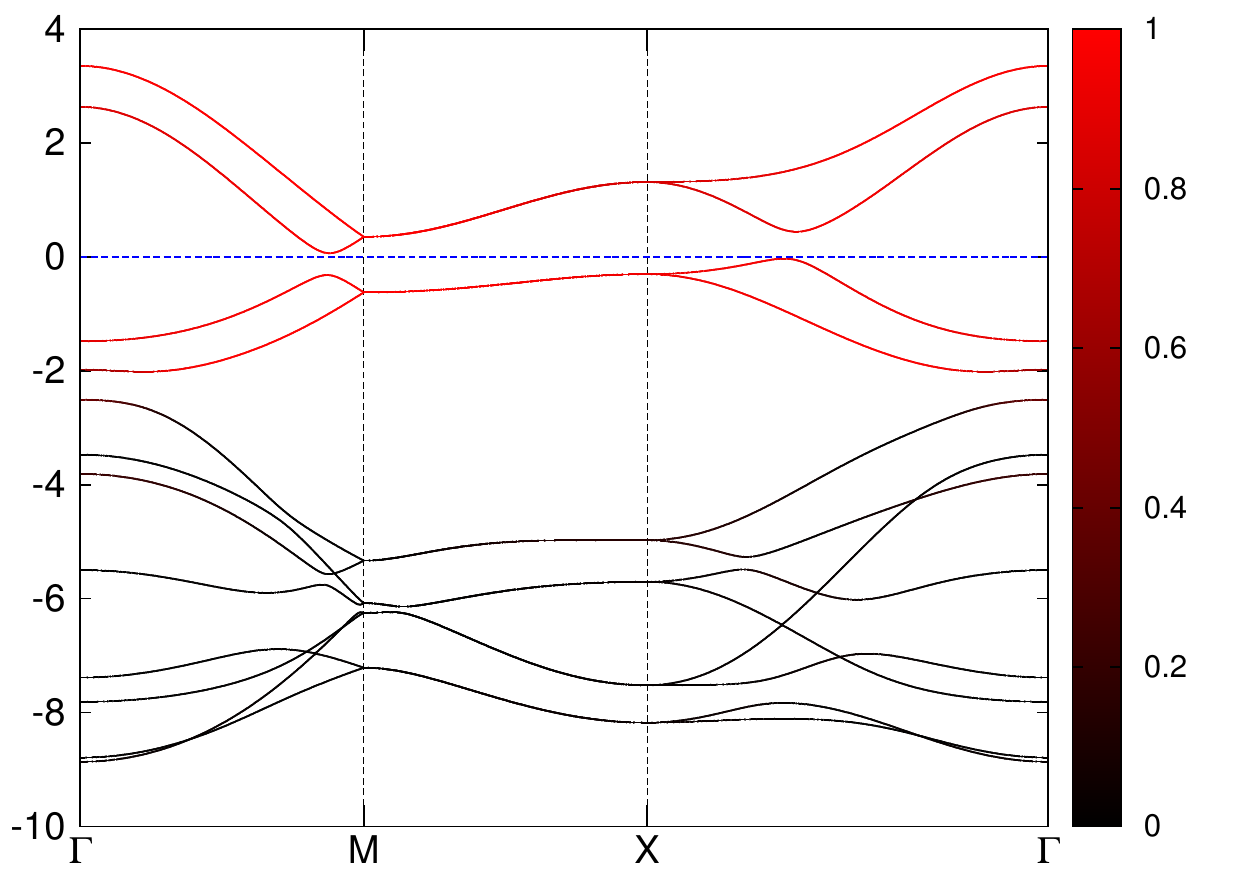}}

\endgroup

\begingroup
\label{fig:113-band}
\subfloat[][\label{fig:113-1}$\lambda= 0t$]{     
 \includegraphics[scale=0.43]{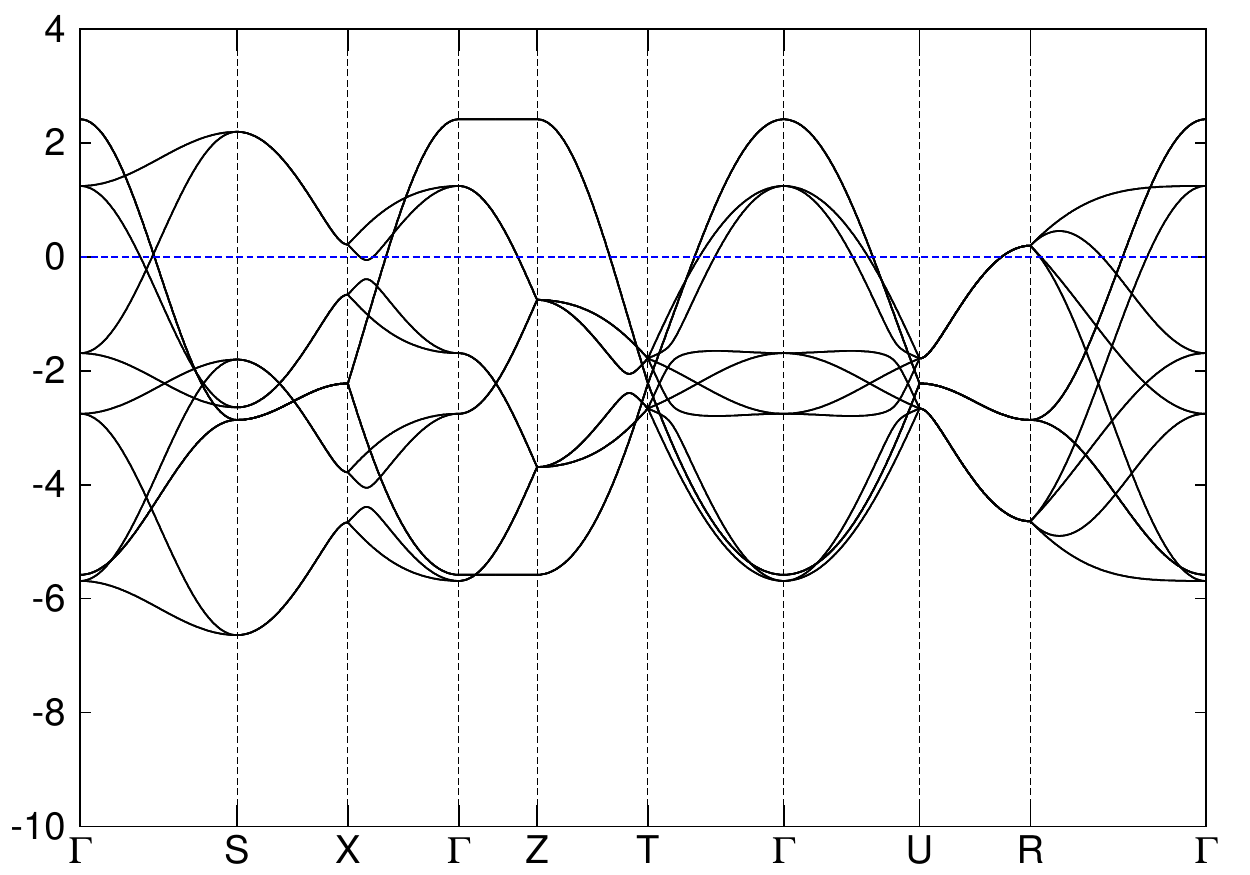}} \quad
\subfloat[][\label{fig:113-bs-second}$\lambda=2t$]{     
 \includegraphics[scale=0.43]{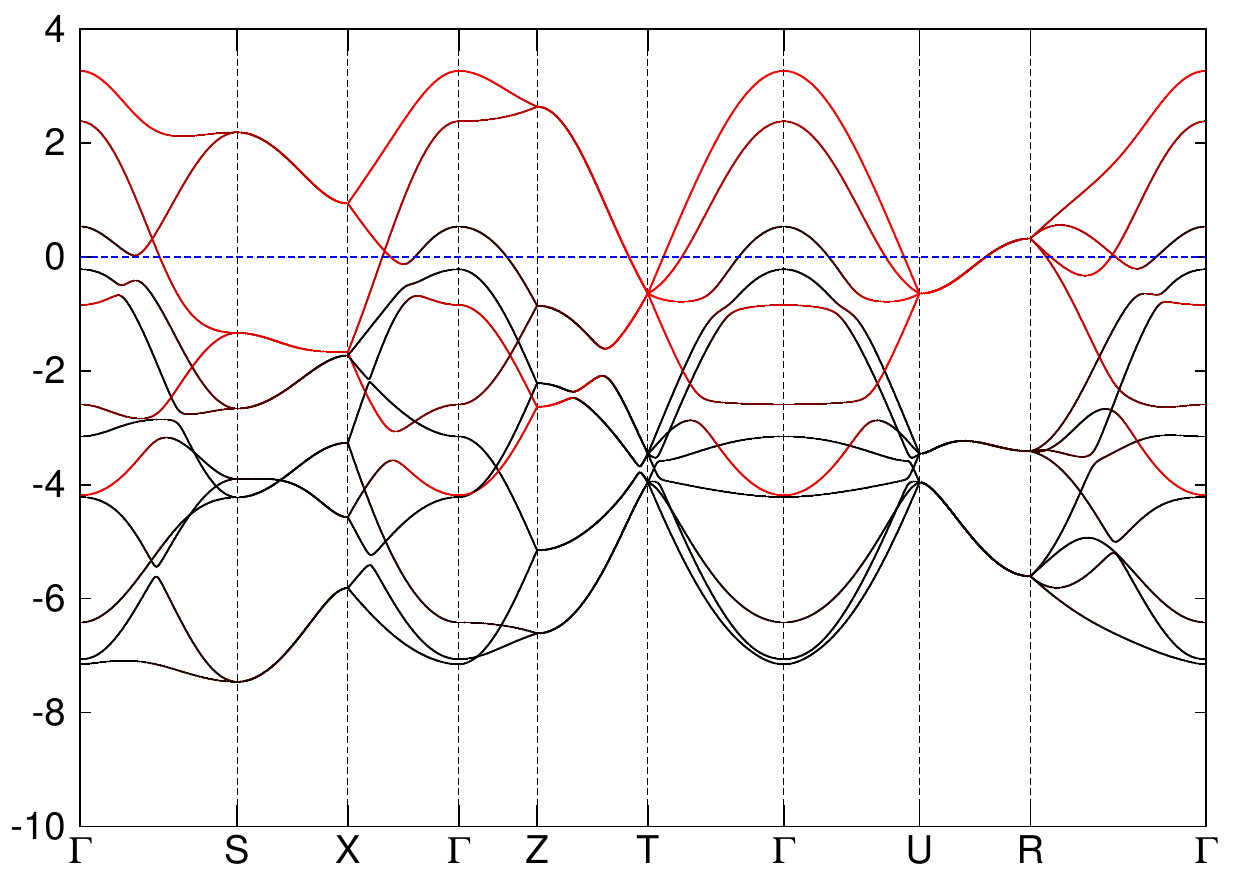}} \quad
\subfloat[][\label{fig:113-3}$\lambda=4t$]{     
  \includegraphics[scale=0.43]{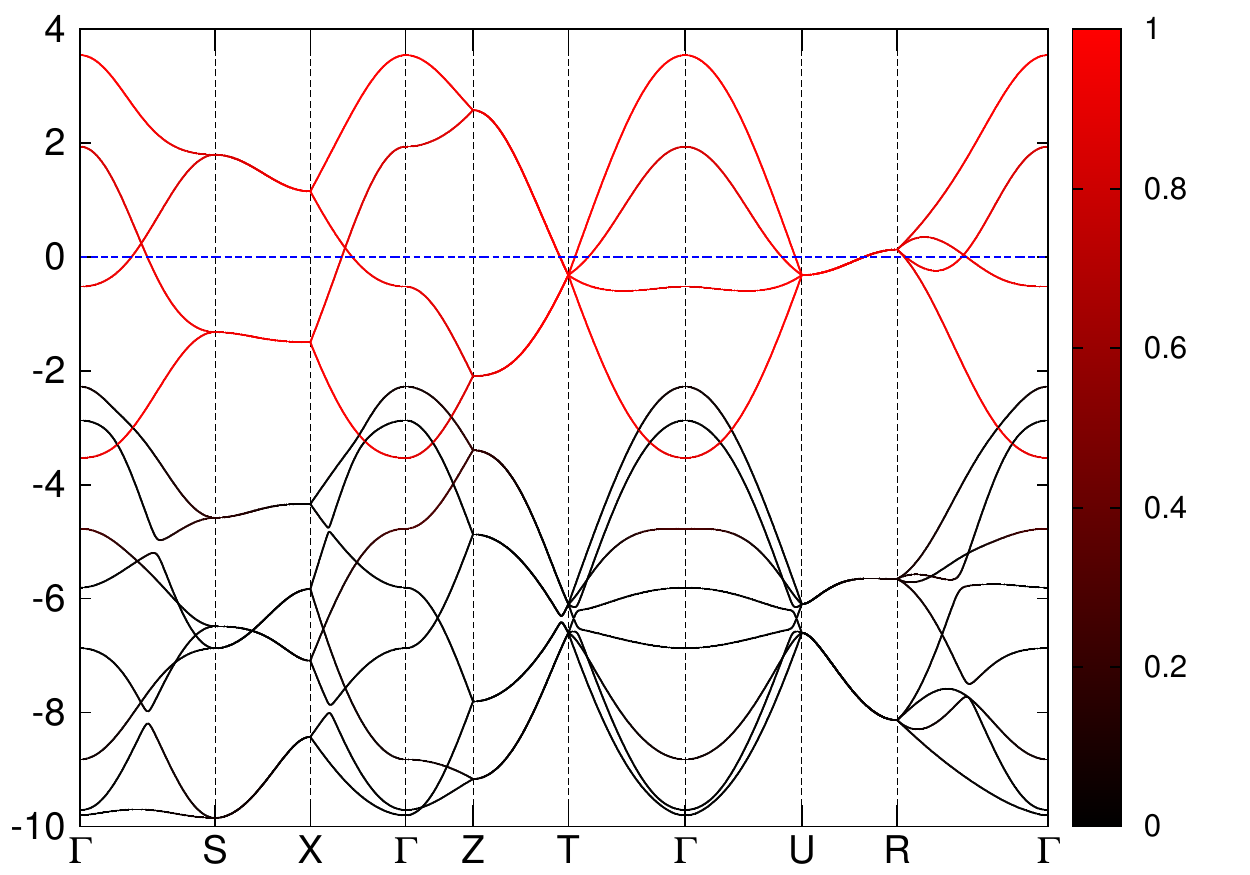}}

\endgroup

\caption{\label{fig:t2g-band}(Color Online) Tight-binding band structure using t$_{2g}$ orbitals for Sr-214 (first row), Sr-327 (second row), Sr-113 (third row) as a function of SOC strength $\lambda$. Red color is used to indicate the weight of J$_{\text{eff}}$=1/2 orbitals in each band and blue horizontal line denotes the Fermi level. For intermediate $\lambda$ (second column), the J$_{\text{eff}}$=1/2 bands are not separated from the J$_{\text{eff}}$=3/2 bands. In Sr-113, the tilting of the oxygen octahedra in addition to rotation leads to four Ir atoms in a unit cell \cite{Carter:2012fk}. Here, to enable comparison, the Brillouin zone is folded with a wavevector Q=$(0,0, \pi)$, but the effect of tilting of the oxygen octahedra on the band degeneracy is not included for simplicity.}
\end{figure*}


{\it Tight-binding model for Sr-214} - Sr-214 is a layered compound with a unit cell that contains four layers. Within each layer, the oxygen octahedra rotate about the c-axis in a staggered fashion contributing two Ir atoms (which we label blue (B) and red (R)) as shown in Fig. \ref{Fig:Mag_ordering}(a) to the unit cell. For the sake of clarity, although the actual unit cell has eight Ir atoms,  
 we focus on a single IrO$_2$ layer with two Ir atoms in this section. The hoppings between the various layers will be discussed later in Sec. \ref{sec:Model}.

Let $d^{\dagger,\gamma,\alpha}_{k,\sigma}$ denote the creation operator for an electron with spin $\sigma = \ua, \da$, orbital $\alpha =yz, xz, xy$ in sublattice $\gamma =$ blue (B), red(R). Using the spinor $\psi_{k} = [d^{B, yz}_{k , \da}, d^{B, xz}_{k , \da},d^{B, xy}_{k , \ua},B \leftrightarrow R, \ua \leftrightarrow \da]^{T}$,  we can write the tight-binding Hamiltonian as
$\sum_k \dagg{\psi}_k H \psi_{k}$. Considering both nearest-neighbour (NN) and next nearest-neighbour (NNN) hoppings, the matrix $H$ in this basis is given by
\begin{align}
  H = \begin{pmatrix}
H_{so} + H_{BB} & H_{BR} \\
\dagg{H}_{BR} & H_{so} + H_{RR} 
\end{pmatrix} + [\uparrow \leftrightarrow \downarrow],
\end{align}
where $H_{so}$, the atomic spin-orbit coupling $\lambda \vect{L}_{i} \cdot \vect{S}_{i}$, is \begin{align}
H_{so} = \begin{pmatrix}
       0& i \lambda/2 & -\lambda/2 \\ 
     -i \lambda/2 & 0 & i \lambda/2 \\
     - \lambda/2 & -i \lambda/2 & 0 
   \end{pmatrix}.
\end{align}

H$_{BR}$ and H$_{BB/RR}$ represent the NN and NNN hopping matrices respectively and are given by
\begin{align}
H_{BR} = \begin{pmatrix}
\epsilon^{yz}_{n} & \epsilon^{rot}  & 0 \\
-\epsilon^{rot} & \epsilon^{xz}_{n} & 0 \\
0 & 0 & \epsilon^{xy}_{n} 
\end{pmatrix}, \,
 H_{BB / RR} = \begin{pmatrix}
\epsilon^{yz}_{d} & \epsilon^{1d}  & 0 \\
\epsilon^{1d} & \epsilon^{xz}_{d} & 0 \\
0 & 0 & \epsilon^{xy}_{d}
\end{pmatrix},
\end{align}
where the dispersions $\epsilon^{yz}_n$, $\epsilon^{xz}_n$,$\epsilon^{rot}$, $\epsilon^{xy}_n$, $\epsilon^{yz}_d$,$\epsilon^{xz}_d$,$\epsilon^{1d}$, and $\epsilon^{xy}_d$ are listed in App. \ref{app:t2g-214}.

{\it Tight-binding model for Sr-327} - Sr-327, like Sr-214, has a layered structure with a unit cell that spans four layers. Unlike Sr-214, Sr-327 has a bilayer structure with two bilayers as shown in Fig. \ref{Fig:Mag_ordering}(b) in the unit cell\cite{Subramanian:1994fk}. Each layer resembles Sr-214 with its staggered oxygen octahedra rotation and the stacking is such that octahedra neighbouring each other in adjacent layers of a bilayer rotate in opposite directions. A single bilayer with four Ir atoms is considered for the $t_{2g}$ tight-binding model.

Following previously employed notation, a spinor $\psi_{k, l} = [d^{l, B, yz}_{k , \da}, d^{l, B, xz}_{k , \da},d^{l,B, xy}_{k , \ua},B \leftrightarrow R, \ua \leftrightarrow \da]^{T}$ where the additional index $l=1,2$ denotes the two different layers in the bilayer is employed to write the tight-binding Hamiltonian as
$\sum_k \dagg{\psi}_{k,l} H^{l l'} \psi_{k, l'}$. Including up to NNN hoppings, the matrix $H$ is given by
\begin{align}
  H^{l l'} = \begin{pmatrix}
H_{so} \delta_{l l'}+ H^{l l'}_{BB} & H^{l l'}_{BR} \\
H^{\dagger ll'}_{BR} & H_{so}\delta_{l l'} + H^{l l'}_{BB} 
\end{pmatrix} + [  \uparrow \leftrightarrow \downarrow],
\end{align}
with $l=l'$ reproducing the Hamiltonian for Sr-214 discussed in the previous section. $l\neq l'$ is the hopping between the layers and consists of two parts
\begin{align}
\label{eq:t2g-bilayer-hopping}
  H^{12}_{BR} = \begin{pmatrix}
t_z & t_z^{\prime} & 0 \\
-t_z^{\prime} & t_z & 0\\
0 & 0 & t_z^{\delta}
\end{pmatrix}; \,  \,
   H^{12}_{BB} = \begin{pmatrix}
\epsilon_b^{yz} & 0 & 0 \\
0 & \epsilon_b^{xz} & 0\\
0 & 0 & 0
\end{pmatrix},
\end{align}
where the first part $H^{12}_{BR}$ includes hopping from a blue atom to a red atom immediately on top of it and the second part $H^{12}_{BB}$ denotes hopping from a blue atom to the four nearest blue atoms on the adjacent layer. In the first part, $t_z$, $t^{\prime}_z$, and $t^{\delta}_z$ represent the intra-orbital hopping for $d_{yz/xz}$ orbitals, hopping from a $d_{yz}$ orbital to a $d_{xz}$ orbital, and hopping between $d_{xy}$ orbitals respectively. The dispersions $\epsilon^{yz}_{b}$, $\epsilon^{xz}_{b}$ in $H^{12}_{BB}$ are described in App. \ref{app:t2g-327}.

\label{t2g:113}
{\it Tight-binding model for Sr-113} - Sr-113 is a three-dimensional compound which crystallizes in an orthorhombic perovksite structure under pressure \cite{Longo:1971bh, Zhao:2008nx}. The oxygen octahedra surrounding the Ir atoms are rotated about the z-axis and tilted about the [110] direction in such a way that there four different Ir atoms in the unit cell. Here, for the sake of simplicity \footnote{This has been treated in Ref. \onlinecite{Carter:2012fk}}, we will not include the tilting of the octahedra along the [110] direction. In the same spinor basis used for Sr-214, the tight-binding Hamiltonian is
\begin{align}
  H = \begin{pmatrix}
H_{so} + H_{BB} +H_z& H_{BR} \\
\dagg{H}_{BR} & H_{so} + H_{RR} + H_z 
\end{pmatrix} + [ \uparrow \leftrightarrow \downarrow],
\end{align}
with $H_{BB}$ and $H_{RR}$ identical to Sr-214 and the additional matrix $H_z$ denoting c-axis dispersions given by 
\begin{align}
  H_z = \begin{pmatrix}
\epsilon_{z} & 0  & 0 \\
0 & \epsilon_{z} & 0 \\
0 & 0 & \epsilon_z^{\delta}
\end{pmatrix}.
\end{align}
Here, $\epsilon_z$ and $\epsilon_z^{\delta}$ denote intra-orbital hopping along the c-axis for $d_{yz/xz}$ and  $d_{xy}$ orbitals respectively are shown in App. \ref{app:t2g-113}.

The band structures for Sr-214, Sr-327, and Sr-113 are plotted in Fig. \ref{fig:t2g-band}. The three rows in the figure represent Sr-214, Sr-327 and Sr-113, respectively.
The three columns represent no SOC, intermediate SOC and large SOC, respectively. The J$_{\text{eff}}$=1/2 bands in columns 2 and 3 have been colored red. When SOC is absent (column 1), increasing the dimensionality does not have significant effect on the bandwidth of the $t_{2g}$ orbitals as mentioned in the introduction. For spin-orbit coupling large enough to separate the J$_{\text{eff}}$=1/2 orbitals from the J$_{\text{eff}}$=3/2 orbitals (column 3), the bandwidth does increase with dimensionality. With intermediate SOC (column 2), the occupied states are a mixture of J$_{\text{eff}}$=1/2 and J$_{\text{eff}}$=3/2, and the J$_{\text{eff}}$=1/2 bandwidth W$_j$ is not well-defined for this case. From the colors, we can infer that while there is significant mixing between J$_{\text{eff}}$=1/2 and J$_{\text{eff}}$=3/2, the bands near the Fermi level are mostly comprised of J$_{\text{eff}}=1/2$ orbitals. Note that SOC in combination with the bilayer hopping makes Sr-327 almost insulating with a direct gap at every $k$ point. This intermediate coupling band structure is also consistent with those seen in LDA+SO calculations. Having identified the bands near the Fermi level, an effective tight-binding model for the three compounds using J$_{\text{eff}}$=1/2 orbitals is derived below.   

\section{Tight-binding model in \texorpdfstring{J$_{\text{eff}}=\frac{1}{2}$}{J=1/2} basis}
\label{sec:Model}

\begin{figure}
\subfloat[][\label{Fig:214_bs}Sr-214]{     
 \includegraphics[scale=0.3]{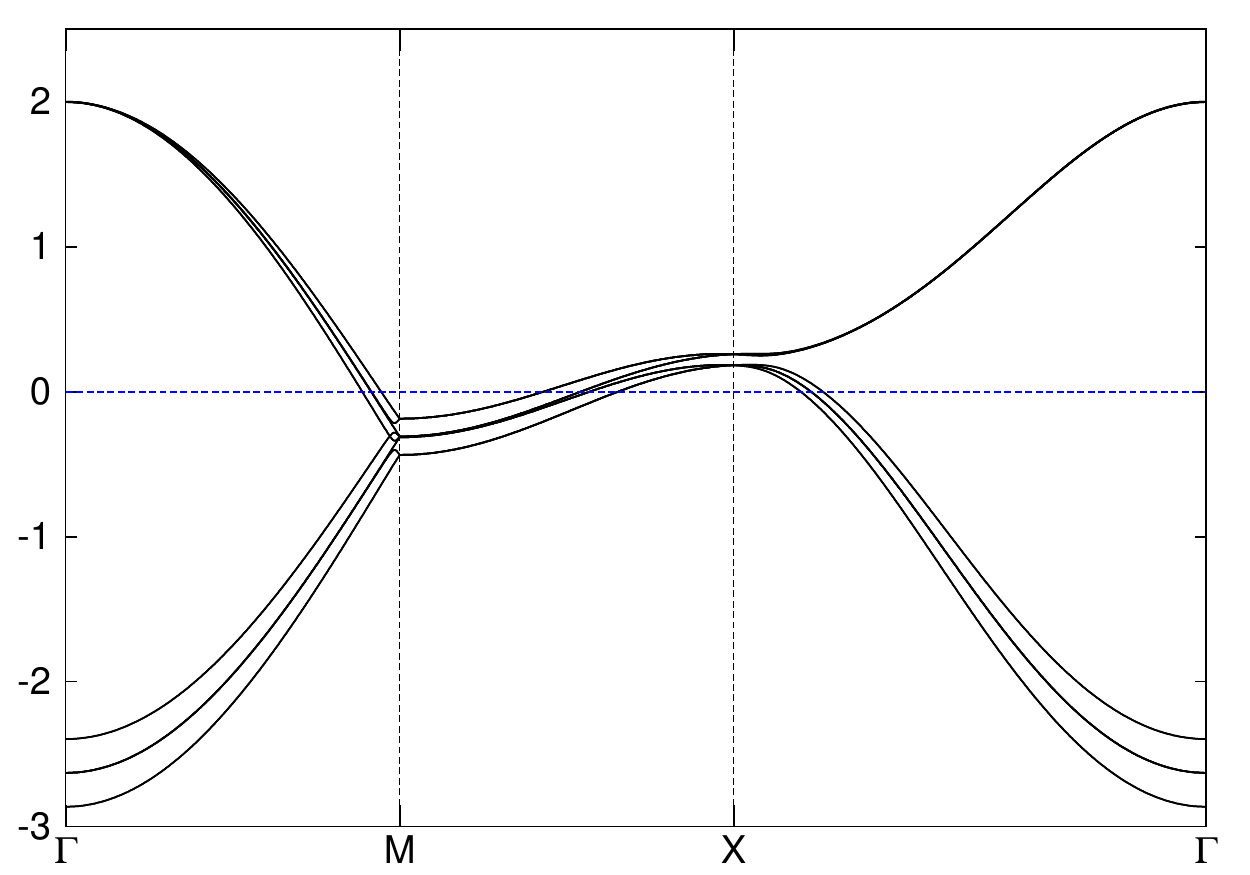}}\,  
\subfloat[][Sr-327]{\includegraphics[scale=0.3]{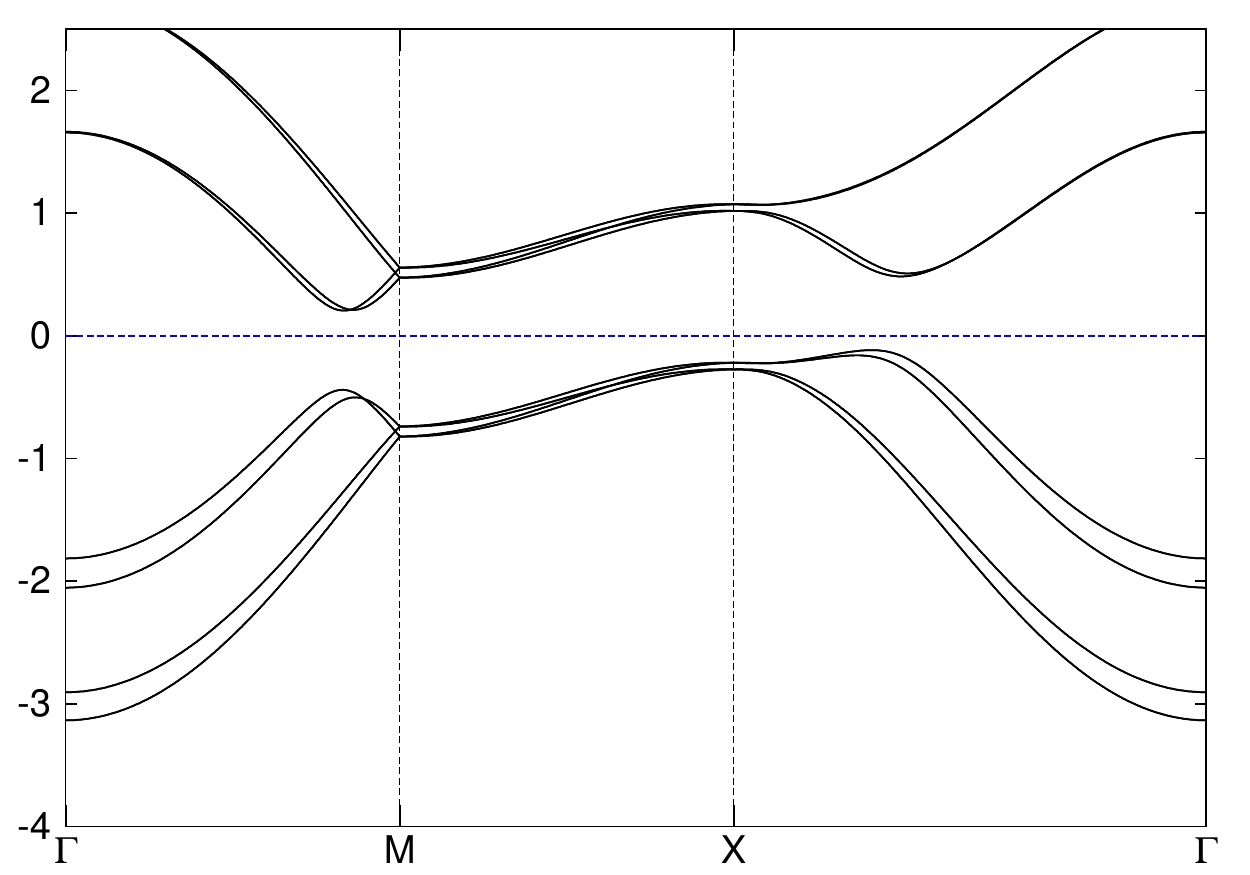}
 \label{Fig:327_bs}}
\caption{Band structure for the J$_{\text{eff}}=1/2$ tight-binding model for (a) Sr-214 and (b) Sr-327. At the Fermi level, this is consistent with LDA calculation and $t_{2g}$ tight-binding model, shown in the third column of Fig. \ref{fig:t2g-band}. In contrast to Fig. \ref{fig:t2g-band}, the full unit cell including the four layers for Sr-214 and the two bilayers for Sr-327 is considered here.} 
\end{figure}
Within weak-coupling theory, bands near the Fermi level rather than all bands are relevant, and the states near the Fermi-level are made up of mostly J$_{\text{eff}}$=1/2 orbitals as shown in the previous section. Thus an effective Hamiltonian for the J$_{\text{eff}}=1/2$ orbitals is obtained by changing to the angular momentum $J$ basis and projecting out states not belonging to J$_{\text{eff}}$=1/2. The J$_{\text{eff}}=1/2$ states are given by    
\begin{align}
\ket{J_z=\pm\frac{1}{2}} = \frac{1}{\sqrt{3}}(\ket{d_{yz},\mp s} \pm i\ket{d_{xz},\mp s} \pm \ket{d_{xy},\pm s}).
\end{align}

In the J$_{\text{eff}}=1/2$ basis, a time-reversal invariant hopping Hamiltonian for a bond is of the form
\begin{align}
t \sigma_0 + i \sum_{\alpha =x , y, z}t^\alpha \sigma^\alpha, 
\end{align}
 where the $\sigma$ matrices represent the pseudospin degree of freedom. To a first approximation, the parameters $t$ and $t^\alpha$ can be derived from the $t_{2g}$ hoppings as done in App. \ref{app:t2gtoj}.  

\label{sub:sr__2_iro__4_model}
As previously mentioned, the unit cell of Sr-214 comprises of four distinct layers with two sublattices in each layer. Here, we include all the layers and use a spinor basis $\phi_k^l = (c^{lB}_{k +},c^{lB}_{k -},c^{lR}_{k +},c^{lR}_{k -})^{T}$ to represent our Hamiltonian, where $l=1,2,3,4$ denotes the layer, $B,R$ are the two sublattices in each layer, and +, - represent the pseudospin states J$_z = 1/2$ and J$_z = -1/2$ respectively. 

Restricting ourselves to hoppings within a layer and between adjacent layers, the Hamiltonian can be written as $H_0 = \sum_k \phi^{\dagger l}_k H^{l l^{\prime}}_k\phi^{l{^\prime}}_k$, where 
\begin{equation}
H^{l l'}_{k} =
\begin{pmatrix}
	A_k 			& B_k 			& 0 			& e^{ik_z c}C_k^\dagger	\\
	B_k^\dagger & A_k 			& C_k 			& 0 					\\
	0 				& C_k^\dagger	& A_k 			& B_k					\\
	e^{-ik_z c}C_k	& 0 			& B_k^\dagger	& A_k					
\end{pmatrix}.
	\label{eq:SLHamiltonian}
\end{equation}
Here $c$ is the c-axis lattice constant and $A_k$, $B_k$, and $C_k$ are each $4 \times 4$ matrices with $A_k$ having all the intra-layer hoppings, $B_k$ containing inter-layer hoppings between layers 1 and 2 (3 and 4) and $C_k$ containing hoppings between layers 2 and 3 (4 and 1). A single layer model would only have $A_k$. 

Using Pauli matrices $\tau$ to represent the sublattice degree of freedom and $\sigma$ to represent the pseudospin,  the different hopping matrices are given by  
\begin{align}
	A_k &= \epsilon_k^{a} + \epsilon_k^{ad} \tau_x + \epsilon'^{ad}_k \tau_y\sigma_z, \\
	B_{k} &= \epsilon_k^{b} + \epsilon_k^{bd} \tau_x + 
			\epsilon_k^{bz} \tau_y\sigma_z + \epsilon_k^{by} \tau_y\sigma_y + \epsilon_k^{bx} \tau_y\sigma_x, \\
	C_k &= \epsilon_k^{c} + \epsilon_k^{cd} \tau_x + 
			 \epsilon_k^{cz} \tau_y\sigma_z + \epsilon_k^{cy} \tau_y\sigma_y + \epsilon_k^{cx} \tau_y\sigma_x.
\end{align}
where the various dispersions making up $A_k$, $B_k$ and $C_k$ are given in App. \ref{app:jhalf-214}. 

\label{sub:sr__3_ir__2_o__7_model}
The non-interacting Hamiltonian in the J$_{\text{eff}}=\frac{1}{2}$ for Sr-327 is similar to that for Sr-214 in Eq. \ref{eq:SLHamiltonian}, with identical matrices $A_k$ and $C_k$. The bilayer nature of this material makes $B_k$ very different. Instead of weak inter-layer hopping terms, there are large bilayer hoppings of the form $B_k = t_c \tau_x + t'_c\tau_y\sigma_x$, where $t_c$ represents the hopping from a blue/red atom to a red/blue atom with the same J$_z$ and $t^{\prime}_c$ is the hopping from an orbital on a blue/red atom to an orbital on a red/blue atom with different J$_z$. The underlying band structure for this compound is shown in Fig. \ref{Fig:327_bs} and the parameters used are given in App. \ref{app:jhalf-327}. The important difference  between Sr-327 and Sr-214 is that, for any set of parameters, Sr-327 has direct gap at each wave-vector. The difference in the nature of the magnetic transition for the Hubbard model can be traced back to this feature of the band structure.


\label{sub:sriro-3-model}
The J$_{\text{eff}}=1/2$ model for the three-dimensional compound Sr-113 has been described in Ref. \onlinecite{Carter:2012fk}. The J$_{\text{eff}}=1/2$ band structure shown in Ref. \onlinecite{Carter:2012fk} includes tilting about [110] leading to gap opening at some of the band crossing points in Fig.
\ref{fig:113-3}.


\section{Magnetism: Mean-Field theory} 
\label{sec:magnetism-mean-field-theory}
The quasi-2D members of the Ruddelsden-Popper series of strontium iridates display novel J$_{\text{eff}}=1/2$ magnetism. To understand and to account for the differences between the magnetic ordering at low temperatures between the different compounds within a consistent picture, we study the effect of electronic interactions by supplementing our J$_{\text{eff}}=1/2$ tight-binding model with a Hubbard term of the form
\begin{equation}
	H_{int} = \tilde{U}\sum_{i}n_{i +}n_{i -},
\end{equation}
where $n_{i \pm}$ is the number operator for pseudospin J$_z= \pm 1/2$ at site $i$. The interaction strength $\tilde{U}$ in the J$_{\text{eff}} = 1/2$ basis can be obtained from a multi-orbital Hubbard model as $\tilde{U} =\frac{U}{3} + 2 \frac{U^{\prime}}{3} - 2\frac{J}{3}$, where $U$, $U^{\prime}$ and $J$ are respectively the intra-orbital repulsion, the inter-orbital repulsion and the Hund's coupling for $t_{2g}$ orbitals. The Kanamori form \cite{kanamori1963electron} with $J=\frac{U-U^{\prime}}{2}$ and\cite{Arita:PRL}  $U^{\prime}=0.8U$ were used in the rest of the paper.
  
Rewriting $	H_{int}=  - \frac{2 \tilde{U}}{3}(S_{ix}^2 + S_{iy}^2 + S_{iz}^2) + \frac{\tilde{U}}{2}\sum_i(n_{i +}+ n_{i +})$, where $S_{i x}, S_{i y}$, and $S_{iz}$ are the components of the pseudospin operator, it is clear that a mean-field decoupling in a magnetic channel such as $S^2_{i \alpha} = 2 m_{i \alpha}S_{i \alpha} - m^2_{i \alpha}$ with $\alpha = x, y, z$ can be employed. A self-consistent iteration to compute the mean-field parameters $m_{i \alpha}$ is carried out.

\begin{figure}[!ht]
\includegraphics[scale=0.6]{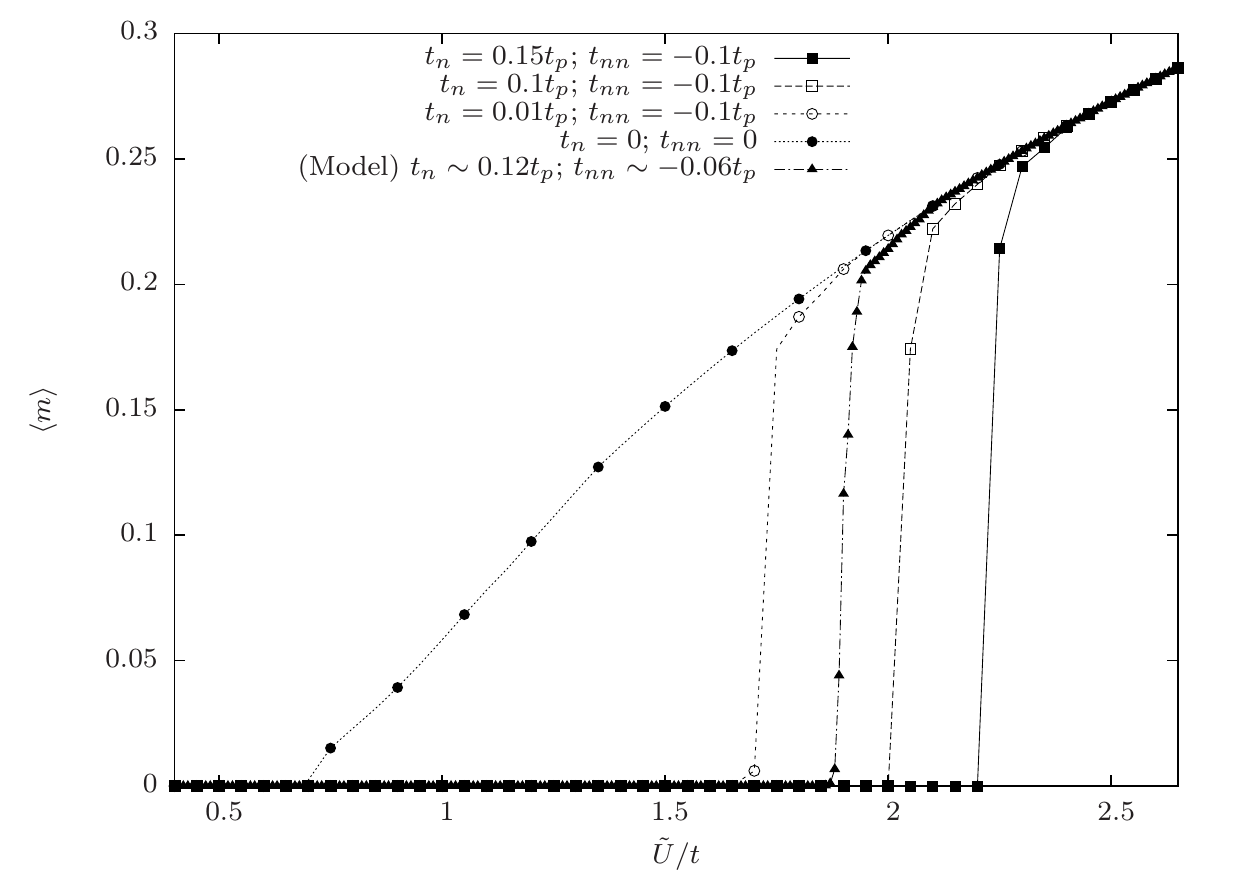}
\caption{Magnitude of order parameter as a function of $\tilde{U}/t$ for Sr-214. 
Various cases of $t_n$ and $t_{nn}$, while keeping other parameters listed in
App. B1, are presented.   For $t_{nn}=t_n =0$, the transition occurs at $0.7t$ and is second order.
On the other hand, when they are finite, the critical interaction strength is larger and the transition is first-order. }
\label{Fig:214-transition}
\end{figure}

{\it Magnetism in Sr-214} -
\label{sub:sr__2_iro__4_meanfield}
On increasing $\tilde{U}$, we find a first order phase transition from a metallic state to a magnetically ordered insulating state as depicted in Fig. \ref{Fig:214-transition}. The ordering pattern, shown in \ref{Fig:Mag_ordering}(a), has the order parameters $\vec{m}_B^1 = (-m_x, m_y, 0)$ for the blue sublattice in the first layer, and $\vec{m}_R^1 = (m_y, -m_x, 0)$ for the red sublattice in the first layer. The ordering in the second layer is identical to the first, while the order parameter for the respective sublattices in the third and fourth layer are opposite in sign from those in the first and second layer.


Within the plane, the canting is such that the net ferromagnetic moment points in the [110] direction. Along the c-axis, the net ferromagnetic moment shows an ``up-up-down-down'' pattern which is consistent with resonant X-ray scattering experiments \cite{JunghoKim} and remains unchanged upon taking the the large $\tilde{U}$ limit of our calculation. 


For the set of parameters mentioned in App. \ref{app:jhalf-214}, the critical interaction strength at which the magnetic transition occurs is $\tilde{U}_c \approx 1.91t$, where $t$ is the magnitude of the Slater-Koster\cite{Slater:1954tg} hopping $t_{dd\pi}$. This critical interaction strength is quite sensitive to the strength of the next-nearest and next-next nearest neighbour hopping, which control the dispersion along the high symmetry direction M-X. In the limit where these hoppings are absent, the transition is second order.  The critical interaction strength also decreases in this limit because of the large Fermi-surface nesting. By varying the strength of the next-nearest neighbour hopping strength between 0 and 0.17, the critical interaction strength can be tuned from $\tilde{U}_c \approx 0.7t$ and $\tilde{U}_c \approx 2.2t$ as shown in Fig. \ref{Fig:214-transition}. Regardless of when the transition occurs, the size of the order parameters for a given $\tilde{U} > \tilde{U}_c$ (for fixed $\lambda$) are the same, implying that while the exact critical interaction $\tilde{U}_c$ hinges on the details of the band structure, the size of the order parameter only depends on $\tilde{U}$ within the J$_{\text{eff}}=1/2$ model. It was shown that SOC and Hubbard U change the order parameter size in $t_{2g}$ model \cite{Carter:327}.

\begin{figure}[!ht]
\label{Fig:bs-ordered}
\subfloat[][\label{Fig:214-bs-ordered}Sr-214]{\includegraphics[width=1.6in]{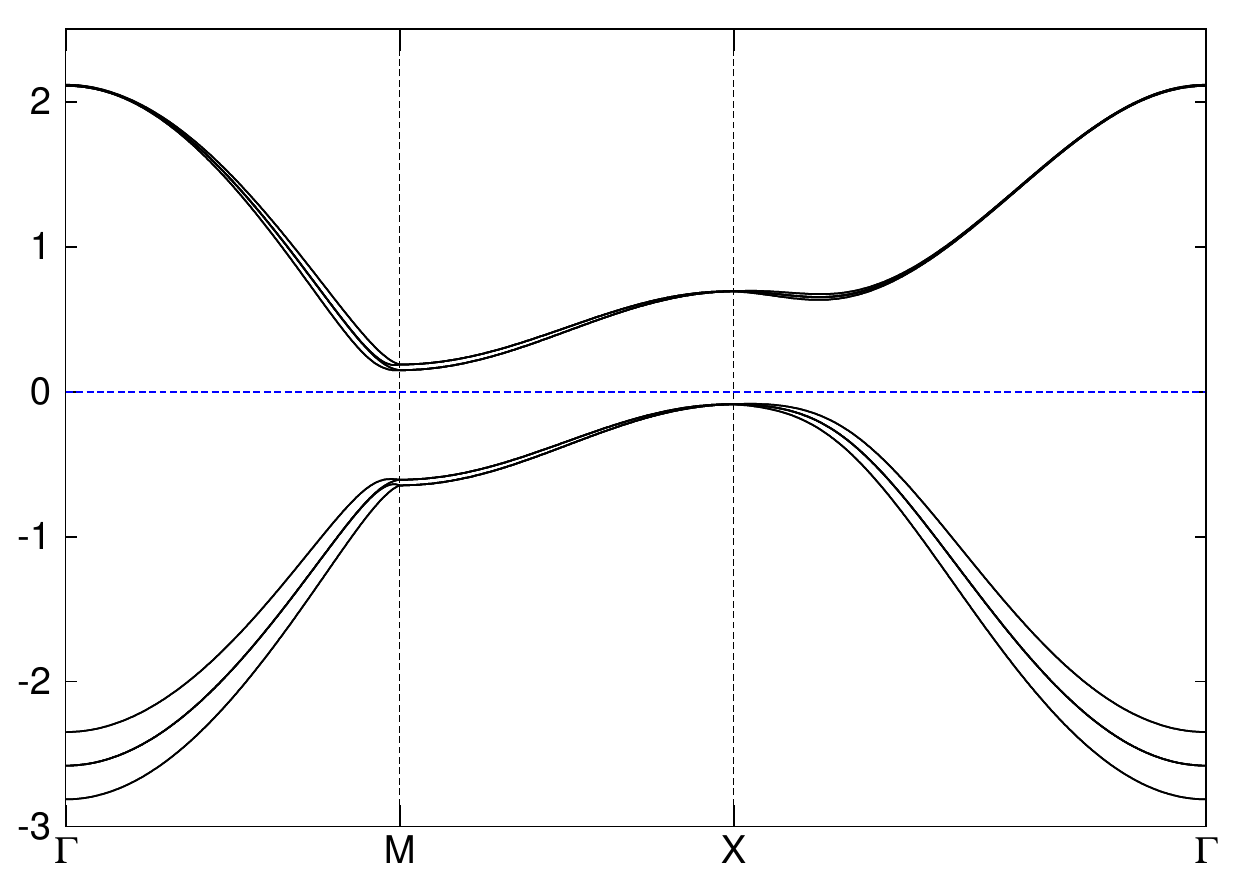}}
\subfloat[][\label{Fig:327--bs-ordered}Sr-327]{\includegraphics[width=1.6in]{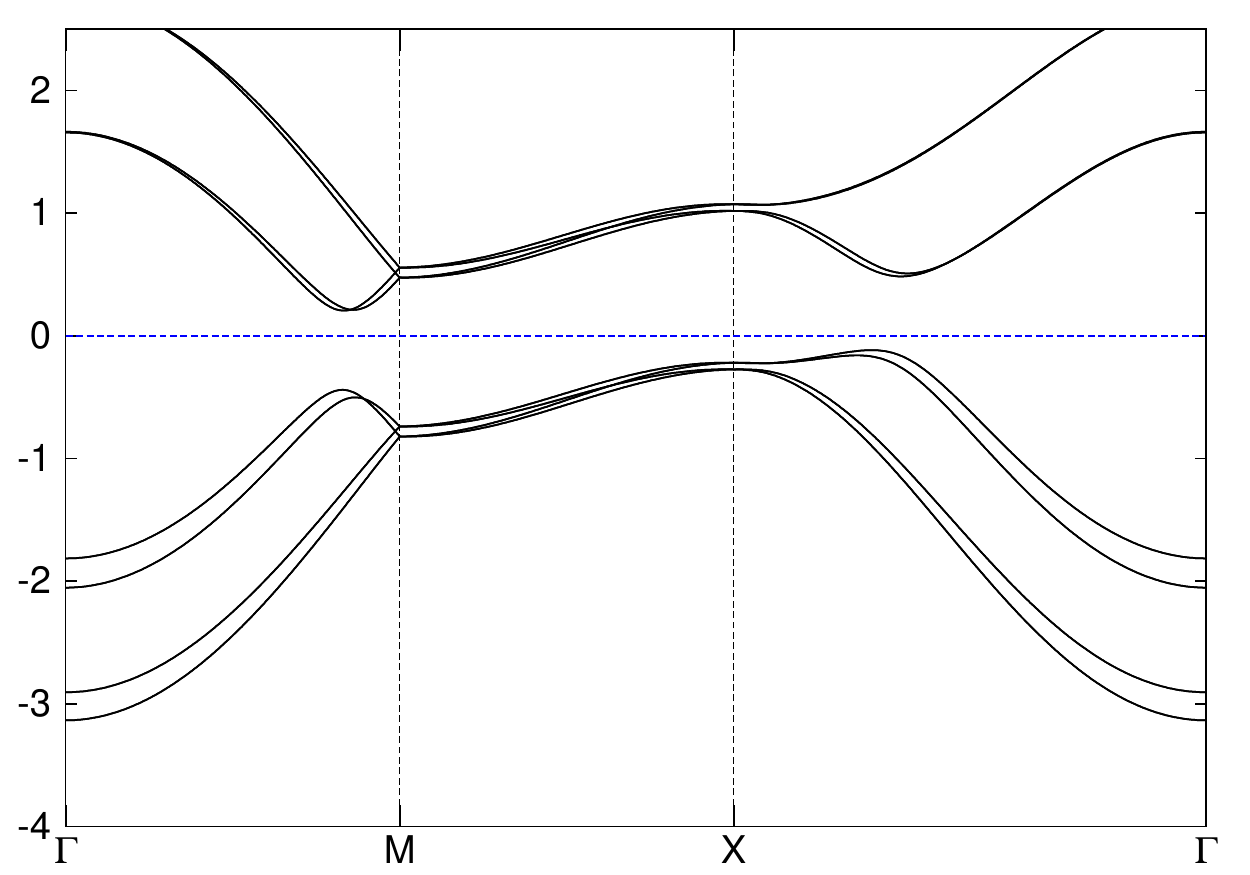}}
\caption{Band structure including Hubbard interaction in the J$_{\text{eff}}=1/2$ basis for Sr-327 and Sr-214. In this figure, a $\tilde{U}$ value of 2.3$t$ was used.}
\end{figure}

The mean-field band structure for $\tilde{U}=2.3t$ is shown in Fig. \ref{Fig:214-bs-ordered}. The charge gap in this model is approximately $0.2t$. We see that ordering opens a gap for moderately large $\tilde{U}$, which in the weak-coupling mean-field treatment hints at Slater insulating behaviour. A Slater transition is usually accompanied by translation symmetry breaking, but for this compound the staggered rotation of the oxygen octahedra has already broken this symmetry. 

{\it Magnetism in Sr-327} -
\label{sub:sr__3_ir__2_o__7_meanfield}
Sr-327, because of its structural similarity to Sr-214, was also expected to have a canted AF ground state. However, the net ferromagnetic moment in Sr-327, was found to be smaller than that of Sr-214\cite{Fujiyama2012Weak, Nagai:2007vn}, although other studies suggest a different moment size \cite{ Dhital:100401}. We show here that the difference in the c-axis stacking, with octahedra in adjacent layers having opposite rotation, makes the c-axis collinear AF state lower in energy than non-collinear state. 

\begin{figure}
\includegraphics[scale=0.6]{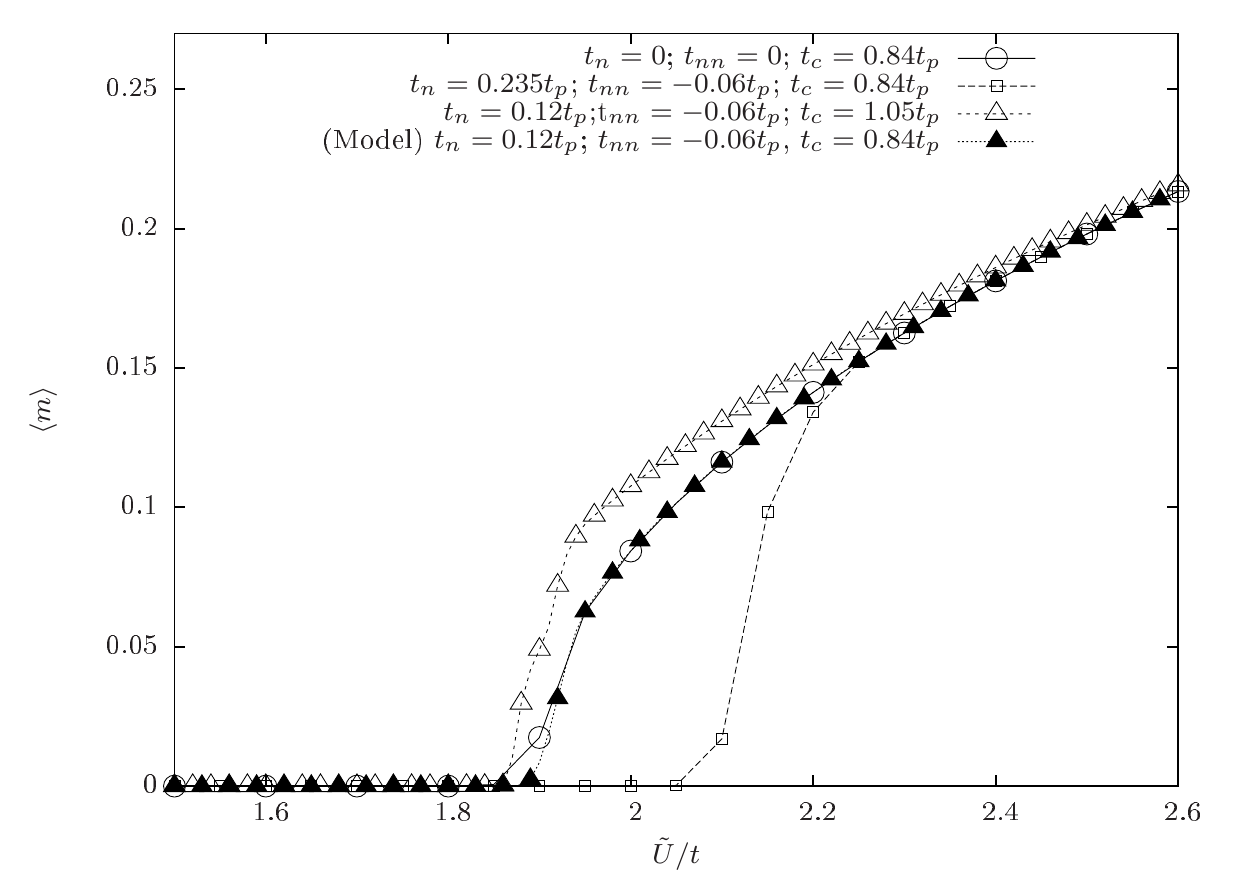}
\caption{Magnitude of order parameter as a function of $\tilde{U}/t$ for Sr-327. Results for different values of $t_n$, $t_{nn}$ and $t_c$,  while keeping the other parameters listed in App. B1, are presented. 
The critical $\tilde{U}_c$ depends on 
mostly $t_n$ and $t_c$, but the size of the moment is insensitive to the tight-binding parameters, once order has set in.} 
\label{Fig:327-transition}
\end{figure}


On increasing $\tilde{U}$, there is a second-order phase transition from a near spin-orbit insulator/metallic state to a magnetically ordered insulator as shown in Fig. \ref{Fig:327-transition} . Unlike the single-layer case, the transition depends more on whether the underlying band structure is gapped or not than on the dispersion in the M-X direction. The ordering pattern, shown in \ref{Fig:Mag_ordering}(b), has the order parameters $\vec{m}_B^1 = (0, 0, m)$ for the blue sublattice in the first layer, and
$\vec{m}_R^1 = (0, 0, -m)$ for the red sublattice in the first layer. The ordering in the second layer is identical to the first, and the order parameter for the respective sublattices in the second bilayer are identical and opposite in sign from those in the first bilayer. 
The ordering pattern can be understood from the terms in the spin model that we describe in Sec. \ref{sub:sr__3_ir__2_o__7_spin}.


The unit cell contains two bilayers and there are two different almost degenerate collinear AF configurations possible for the ground state. We show one (slightly lower in energy) in Fig. \ref{Fig:Mag_ordering}(b), the other configuration has the same spin configuration for both the bilayers. These two configurations were observed in experiments of Ref. \cite{Boseggia:2012}.

For the parameters used here, the critical interaction strength $\tilde{U}_c \approx 1.90t$. This critical interaction strength depends on the actual value of the next-nearest and next-next-nearest neighbour hoppings, similar to the single-layer case, but to a much lesser degree as shown in Fig. \ref{Fig:327-transition}. Magnetic ordering does not change the dispersion but increases the direct gap.    

{\it Magnetism in Sr-113} -
\label{sub:sriro__3_meanfield}
Rotation and tilting of the octahedral environment around the Ir atom contributes more spin-dependent hopping terms in the J$_{\text{eff}} = 1/2$ basis in Sr-113 in comparison to its quasi-2D counterparts. These terms hint at a canted antiferromagnetic ground state with both in-plane and out-of-plane canting. The magnetic transition occuring at an interaction strength $\tilde{U}_c = 2.35t$ takes the system from a semimetal to magnetic semimetal and then to a magnetic insulator with the magnetic pattern $\vec{m}_B = (m_x,m_y,m_z)$ for the Blue sublattice, $\vec{m}_R = (-m_y,-m_x,m_z)$ for the red sublattice, $\vec{m}_Y = (-m_x,-m_y,m_z)$ for the yellow sublattice and $\vec{m}_G = (m_y,m_x,m_z)$ for the green sublattice. The canting within each plane is equal and opposite for the different layers resulting in zero net in-plane moment, but there is a finite magnetic moment in the $z$ direction.

The progression of the critical interaction strength $\tilde{U}_c$ required for a magnetic transition in the Ruddelsden-Popper series explains the experimentally observed MIT, as the number of layers is increased. 
%
In the current mean-field study, the magnetic ordering pattern does not change on increasing $\tilde{U}$. To understand how our weak-coupling results above are related to the strong coupling approach, spin models in the large $\tilde{U}$ limit for Sr$_2$IrO$_4$ and Sr$_3$Ir$_2$O$_7$ are derived in the next section.


\section{Magnetism: Spin-model} 
\label{sec:magnetism_spin_model}

For a general time-reversal invariant hopping between two sites given by $(t_{l l^{\prime}} + i \sum_{\substack{\alpha = x, y,z}}\sigma^{\alpha} t^{\alpha}_{l l^{\prime}}) \dagg{c}_{i, l, \sigma} c_{j, l^{\prime} \sigma^{\prime}}$, the exchange Hamiltonian within second order perturbation theory is 

\begin{align}
	H = 
	\sum_{\substack{<ij> \\  l l^{\prime}}} \left(
	J^{l l^{\prime}}_{ij} \vec{S}_{i}^{l } \cdot\vec{S}_{j}^{l^{\prime}}  + 
	\vec{D}^{l,l^{\prime}}_{ij}  \cdot \vec{S}^{l}_i \times \vec{S}^{l^{\prime}}_j + {\vec S}^{l}_i  \cdot \Gamma^{l,l^{\prime}}_{ij}
	\cdot {\vec S}_j^{l^{\prime}}
	\right),
\end{align}
with Heisenberg coupling $J_{ij}^{l, l^{\prime}} = \frac{4}{\tilde{U}}(t_{l,l^{\prime}}^2 - v_{l,l^{\prime}}^2)$, Dzyaloshinskii-Moriya (DM) coupling $\vec{D}^{l, l^{\prime}}_{ij} =  \frac{4}{\tilde{U}} (2 t_{l, l^{\prime}}\vec{v}_{l, l^{\prime}})$, and anisotropic exchange coupling $\Gamma^{l, l^{\prime}}_{ij, \alpha \beta} = \frac{4}{\tilde{U}}(2 v^{\alpha}_{l, l^{\prime}} v^{\beta}_{l, l^{\prime}})$. Here $\alpha$, $\beta$ are spin components and $l, l^{\prime}$ are layer indices.   

\label{sub:sr__2_iro__4_spin}
For Sr-214, the nearest-neighbour exchange couplings within a layer are $J^{l l}_p = \frac{4}{\tilde{U}}(t_p^2 - t^{\prime 2}_p)$, $\vec{D}^{l l}_p =  \frac{8 \epsilon_i t_pt^{\prime}_p}{\tilde{U}}\hat{z}$, and $\Gamma^{l l}_{p, zz} = \frac{8}{\tilde{U}}t^{\prime 2}_p$ for $l = 1, \ldots, 4$, where $\epsilon_i$ in $\vec{D}^{l l}_p$ represents a change of sign between two adjacent bonds. Ignoring inter-layer couplings, there is a canting angle at which the canted AF state is degenerate to the collinear AF state. This degeneracy arises because it is possible to turn the spin-model into an isotropic Heisenberg model by an appropriate gauge transformation for the spin operators, as has been pointed out by others\cite{Jackeli:2009qf,Wang-2011-Twisted}.  

We now add the inter-layer couplings as a perturbation on the two degenerate states. Since each Ir atom has 4 neighbours in each adjacent layer, we separate the exchange terms to two types, between atoms of the same colour (i.e. B to B) and between atoms of different colours (R to B). There are two hoppings of each kind, which we add together. For Sr-214, between layers 1 and 2, the exchange terms are $J^{12}_{BB} = \frac{4}{\tilde{U}}t_i^2$, $J^{12}_{BR} = \frac{4}{\tilde{U}}(t_{id}^2-t_{iz}^2 -t_{iy}^2-t_{ix}^2)$, $\vec{D}^{12}_{BR} =  \frac{8}{\tilde{U}} (\pm t_{id}t_{ix},\pm t_{id}t_{iy},t_{id}t_{iz})$, and  
\begin{align*}
	\Gamma_{BR}^{12} &= \frac{8}{\tilde{U}}
	\begin{pmatrix}
		t_{ix}^2			&	t_{ix}t_{iy}		&	\pm t_{ix}t_{iz} \\
		t_{ix}t_{iy}		&	t_{iy}^2			&	\pm t_{iy}t_{iz} \\
		\pm t_{ix}t_{iz}	&	\pm t_{iy}t_{iz}	&	t_{iz}^2
	\end{pmatrix}.
\end{align*}
Here, $t_i$ is the inter-layer hopping between B and B (or R and R), $t_{id}$ between B and R, and pseudospin $\sigma_a$ ($a=x,y,z$) dependent hopping between B and R is given by $t_{ia}$. The $\sigma_x$ and $\sigma_y$ dependent hopping between B (R) in layer 1 and the two R(B) in layer 2 have different signs, which we denote by $\pm$. This sign change means that these terms do not contribute to the energy if the magnetic unit cell is the same as the lattice unit cell. For layers 2 and 3, the exchange couplings are identical except a change in sign for the $y$ component of the DM coupling ($D^{23, y}_{BR} = -D^{12, y}_{BR}$) and the $xy$ and $yz$ components of the anisotropic exchange ($\Gamma_{BR}^{23,xy} = -\Gamma_{BR}^{12,xy}$ and $\Gamma_{BR}^{23,yz} = -\Gamma_{BR}^{12,yz}$ ). 

The lattice structure with four neighbours on the adjacent layer for each atom frustrates the DM and isotropic couplings. The canted AF state in an up-up-down-down fashion is favoured by the anisotropic exchange, and this gets picked as the ground state, because the other terms are frustrated by the lattice. 

\label{sub:sr__3_ir__2_o__7_spin}
In Sr-327, the bilayer couplings are much larger than the inter-bilayer couplings, hence only one bilayer is considered. The in-plane exchange couplings are identical to those of the single layer and the out-of-plane couplings are given by $J_c = \frac{4}{\tilde{U}}(t_z^2 - t^{\prime 2}_z)$, $\vec{D}_c = \epsilon_i \frac{8}{\tilde{U}}t_zt'_z\hat{z}$, and $\Gamma^{zz}_c = \frac{8}{\tilde{U}}t^{\prime 2}_z$, where $t_z$, and $t_z^{\prime}$ are respectively the pseudo-spin independent and pseudo-spin dependent bilayer hoppings described in App. \ref{app:jhalf-327}. Unless $\frac{t'_z}{t_z} = \frac{t'_p}{t_p}$, the in-plane and out-of-plane DM couplings are frustrated, i.e. there is no spin configuration such that optimal canting is achieved for both of them. Thus, the relative strength of the bilayer hopping to the in-plane hopping decides the competition between canted AF and collinear AF. From the Slater-Koster method,\cite{Slater:1954tg} we estimate that $\frac{t'_z}{t_z} \approx 3 \frac{t'_p}{t_p}$, picking out the c-axis collinear AF state as the ground state, which has been recently confirmed in experiments \cite{Kim2012Dim, Kim2012Giant, Boseggia:2012, Clancy:2012_arxiv2}.

The inter-layer couplings favour an anti-ferromagnetic pattern between the same atoms of layers 2 and 3 as shown in Fig. \ref{Fig:Mag_ordering}, but this is close in energy to the ferromagnetic pattern, leading to the two possible arrangements discussed previously.

\section{Discussion and Summary} 
\label{sec:discussion_and_summary}
\begin{figure}[]
\begin{overpic}[scale=0.45]{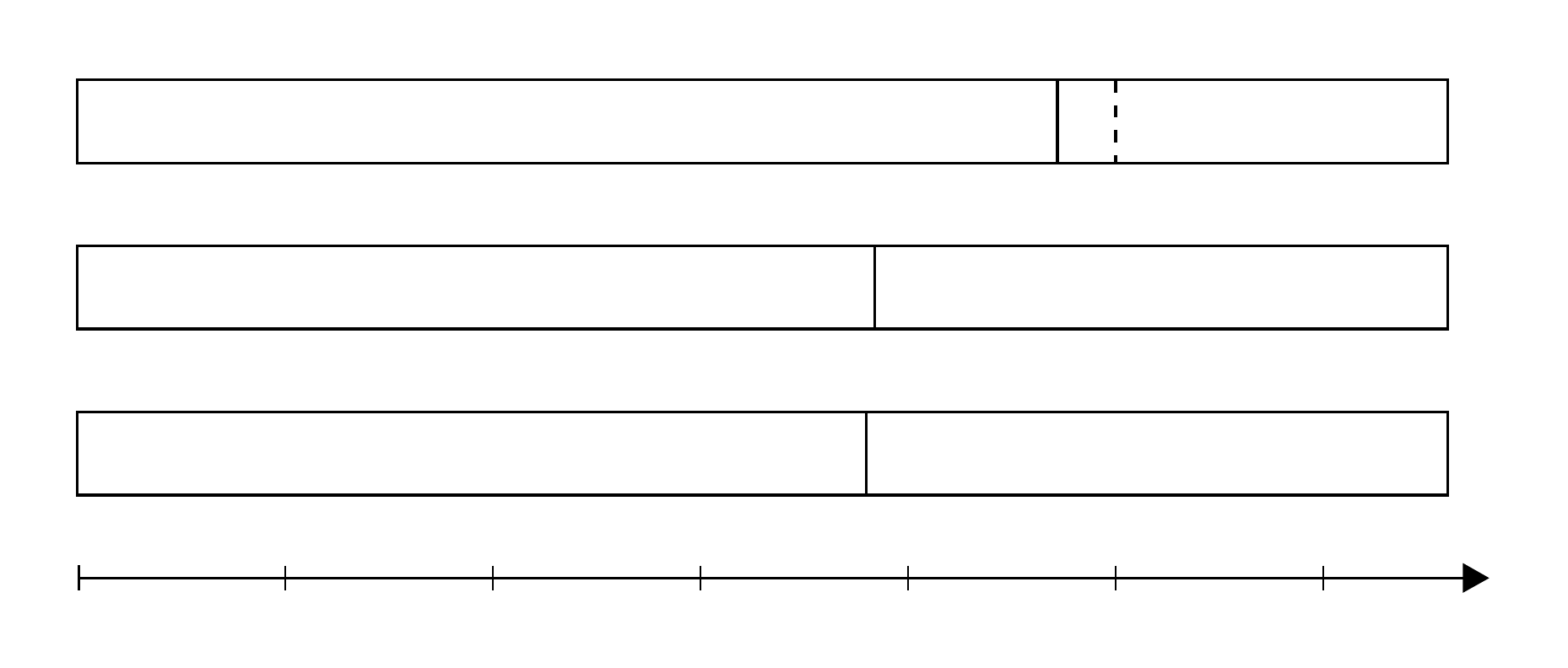}
\put(3,3){{\tiny 0.0}}
\put(16.3,3){\tiny 0.5}
\put(29.6,3){\tiny 1.0}
\put(42.9,3){\tiny 1.5}
\put(56.2,3){\tiny 2.0}
\put(69.5,3){\tiny 2.5}
\put(83,3){\tiny 3.0}
\put(49,0){\scriptsize $\tilde{U}/t$}
\put(-4, 13){\tiny Sr-214}
\put(8, 13){ \textbf{\color{red} \Tiny Paramagnetic Metal}}
\put(65, 13){$\star$}
\put(53.5, 17.5){\scriptsize $U_c$}
\put(69, 13){ \textbf{\color{blue}\Tiny Canted AF Insulator }}
\put(-4, 23.5){\tiny Sr-327}
\put(8,23.5){\textbf{\color{red} \Tiny Spin-orbit band insulator}}
\put(65, 23.5){$\star$}
\put(68, 23.5){\textbf{\color{blue}\Tiny Collinear AF Insulator }}
\put(54, 28.5){\scriptsize $U_c$}
\put(-4, 34){\tiny Sr-113}
\put(8, 34){\textbf{\color{red} \Tiny Paramagnetic semimetal}}
\put(65, 34){$\star$}
\put(66, 39){\scriptsize $U_c$}
\put(68, 33){{\textbf{\color{red} \Tiny M}}}
\put(68, 35.3){\textbf{\color{red}\Tiny M}}
\put(74, 34){\textbf{\color{blue}\Tiny N-CAF }}
\put(82,34) {\textbf{\color{blue} \Tiny Insulator}}
\end{overpic}

\caption{\label{Fig:PhaseDiag}(Color online) Phase diagram for the magnetic phases of the single layer, bilayer and three-dimensional compounds with the accompanying MIT. The critical value of $\tilde{U}$ at which the magnetic transition occurs for each compound is $\tilde{U}_c/t = 1.90$, $1.91$ and $2.35$, respectively. For Sr-214, due to the first-order nature of the transition, the system goes directly from metallic to insulating. For Sr-327, the magnetic ordering does not change the Fermi surface topology, but further increases the existing direct band gap. For Sr-113, the transition turns the metal into a magnetic metal (MM) before it develops a gap as the ordering increases. Sr-113 shows non-coplanar canted antiferromagnet (N-CAF) order, with the ferromagnetic moment pointing along the crystal c-axis. In a scenario where the Hubbard interaction is $2.3t \sim 0.46eV$ (setting $t \sim 200 meV$), the stars represent where each compound should be on the phase diagram.}
\end{figure}
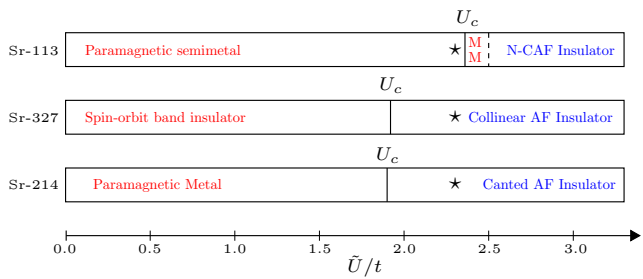

We build a tight-binding model using t$_{2g}$ orbitals for the different layered perovskite iridates in Sec. \ref{sec:t2g} to identify the states near the Fermi level to be made up of mostly J$_{\text{eff}}=1/2$. We then construct a J$_{\text{eff}}=1/2$ tight-binding model in Sec. \ref{sec:Model}, and study the magnetic order and MIT by taking into account Hubbard interactions within a mean-field approximation in Sec. \ref{sec:magnetism-mean-field-theory}. We find that our approach captures a rich phase diagram as $\tilde{U}$ increases; first-order transition from metal to magnetic insulator for Sr-214, transition from a nearly band insulator to magnetic insulator for Sr-327, and semimetal to magnetic insulator via a magnetic semimetal for Sr-113. The phase diagram for the different layered cases, sketched in Fig. \ref{Fig:PhaseDiag}, shows that the different members of the Ruddelsden-Popper series exhibit different critical interaction $U_c$ for a magnetic transition.

The \emph{magnetic ordering pattern} in the ground state is intimately connected to the lattice structure. In Sr-214, the interlayer isotropic Heisenberg and DM interactions are frustrated, because Ir atoms in adjacent layers create a zero effective field due to their positions. Combined with the anisotropic exchange, this frustration selects the ``up-up-down-down'' pattern for the net ferromagnetic moment. Although these inter-layer exchanges are small compared to the intra-layer exchanges, they are significant because in their absence the coplanar CAF and collinear AF are degenerate within our J$_{\text{eff}}=1/2$ model. In Sr-327, in contrast to Sr-214, the collinear AF state along the c-axis is preferred, because the bilayer nature of the lattice frustrates the DM interaction of each Ir atom with its five neighbours. However, this configuration is proximate in energy to the canted AF state and spin dynamics as function of temperature and magnetic field offer interesting subjects for future study.

For the iridates studied here, magnetic ordering patterns are identical to the ones given by the large $\tilde{U}$ limit spin model, implying that their Mott character cannot be corroborated by the ordering pattern. The proximity to a metallic state hints at Slater insulating behaviour. Although a Slater transition is usually concordant with translational symmetry breaking, translational symmetry has already been broken before the onset of magnetic order in Sr-214, due to the staggered rotation of the oxygen octahedra. Sr-327, on the other hand, is either a small band gap insulator or a metal with small Fermi surfaces at $\tilde{U}$ $\approx$ 0.46eV suggesting that the magnetic ordering is not due to Mott physics. ARPES studies above and below the transition temperature might be a way to test this proposal.

The comparison of iridates to the cuprates deserves some discussion. The J$_{\text{eff}} = 1/2$ spin model resembles that of the high-Tc cuprates, and it was proposed\cite{Wang-2011-Twisted} that iridates might exhibit high-Tc d-wave superconductivity on electron doping. However, unlike the two-dimensional $d_{x^2 - y^2}$ orbitals in the cuprates, the J$_{\text{eff}} = 1/2$ state is an equal mixture of the $t_{2g}$ orbitals resulting in an isotropic s-wave like wavefunction. The strong bilayer coupling in Sr-327 which makes this compound a spin-orbit band insulator, unlike the bilayer cuprates where bilayer coupling can be treated perturbatively, highlights this difference. For Sr-214, anisotropic exchange is also not negligible and thus it is unlikely that these iridates will show d-wave singlet pairing, although superconductivity remains an open possibility.

In conclusion, we have studied the metal-insulator transition in the layered iridates within a Slater picture using mean-field theory. Our results suggest that magnetic ordering in the Ruddelsden-Popper series of strontium iridates is explained within weak-coupling theory. Magnetic ordering patterns obtained in the weak-coupling approach are continuously connected to the ordering patterns obtained in large-$\tilde{U}$ spin-model indicating that magnetic ordering is not sufficient to justify the Mott physics in this series of iridates. Further studies on the relation between charge gap and magnetic ordering are necessary to pin down the strong correlation effect in these iridates. 

\textit{Acknowledgements -}
This work was supported by NSERC of Canada. HYK acknowledges the Aspen Center for Physics
where a part of the work was carried out.

\appendix
\section{\label{app:t2g}Dispersions and tight-binding parameters in t$_{2g}$ basis}

 The hopping parameters and dispersions for the $t_{2g}$ tight-binding model for Sr-214, Sr-327 and Sr-113 mentioned in Sec. \ref{sec:214-t2g} are elaborated upon here.

\label{app:t2g-214}
{\it Sr-214 -} - The nearest-neighbour dispersions contained in H$_{BB}$ are 
\begin{align}
\epsilon^{xy}_{n} &= 2 t (\cos(k_x)+\cos(k_y))    \\
\epsilon^{yz}_{n} &= 2 (t_1 \cos(k_y)+ t^{\delta} \cos(k_x)) \\    
\epsilon^{xz}_{n} &= 2 (t_1 \cos(k_x)+ t^{\delta} \cos(k_y)) \\    
\epsilon^{rot} &= 2 t^{\prime}(\cos(k_x) + \cos(k_y)),
\end{align}
where the superscripts $xy,xz, yz$ denote the orbitals and $t$ is the direct Slater-Koster overlap $t_{dd\pi}$ between the Iridium $d$ orbitals.
$t_1$ and $t^{\delta}$ denote hopping integrals of $d_{xz} (d_{yz})$ to $d_{xz} (d_{yz})$ along $x(y)$- and $y(x)$-axis, respectively. $\epsilon^{rot}$ arises from the nearest-neighbour overlap $t^{\prime}$ between $d_{yz}$ and $d_{xz}$ due to the rotation of oxygen octahedra. 
The next-nearest neighbour dispersions making up H$_{BB/RR}$ are
\begin{align}
\epsilon^{xy}_{d} &= 4 t_{n} \cos(k_x)\cos(k_y) \\
\epsilon^{yz}_{d} &= \epsilon^{xz} = 4 t_{nd} \cos(k_x)\cos(k_y) \\
\epsilon^{1d} &= 4 t_{1d} \sin(k_x)\sin(k_y).
\end{align}
where $t_n$ and $t_{nd}$ represent intra-orbital hopping between $d_{xy}$ and $d_{xz/yz}$ respectively. $\epsilon^{1d}$ is hopping between $d_{yz}(d_{xz}) $ and $d_{xz}(d_{yz})$ orbitals parametrized by an overlap $t_{1d}$. All the hopping parameters are obtained from Slater-Koster theory using the overlaps for the $d$ orbitals $(t_{dd \pi}, t_{dd \sigma},t_{dd \delta}) = (-1.0,1.5,0.25) $, a rotation angle of $12^{\circ}$ for the IrO$_6$ octahedra and a suppression factor of 0.2 for next-nearest neighbour overlaps compared to nearest-neighbour overlaps.
The parameters set obtained from Slater-Koster method for the single layer model is then $(t, t_1 , t^{\prime}, t_{n}, t_{1d}, t^{\delta},t_{nd}) =(-1.0, -0.94,-0.15,0.16, 0.11,	 0.27,0.0)$.

{\it Sr-327} -
\label{app:t2g-327}
For Sr-327, within each layer, the dispersions and the tight-binding parameters are identical to those described for Sr-214 in App. \ref{app:t2g-214}. The nearest neighbour overlaps between a blue(B)/red(R) atom and a red(R)/blue(B) atom in Eq. \ref{eq:t2g-bilayer-hopping} were parametrized by $(t_{z}, t^{\prime}_{z} , t^{\delta}_z) = (-0.80, 0.36,-0.15)$. Here $t_z$ is the intra-orbital overlap between $d_{xz/yz}$ orbitals, $t^{\delta}_z$ is between $d_{xy}$ orbitals, and $t^{\prime}_{z}$ is the overlap between a $d_{yz/xz}$ orbital on one atom and $d_{xz/yz}$ orbital on the Ir atom in the next layer. The next nearest neighbour bilayer dispersions making up $H^{12}_{BB}$ in Eq. \ref{eq:t2g-bilayer-hopping} are $\epsilon^{yz}_b = 2 \,t_{zn} \cos(k_y)$ and $\epsilon^{xz}_b = 2 \, t_{zn} \cos(k_x)$ which are orbital diagonal with the parameter $t_{zn} = 0.2$.

{\it Sr-113} -
\label{app:t2g-113}
Since the tilting of the octahedra along the [110] direction was ignored, Sr-113 differs from Sr-214 only in its c-axis dispersions. These intra-orbital hoppings are  
$ \epsilon_z = 2 \,t_{z} \cos(k_z)$ for $d_{yz/xz}$ orbitals and $ \epsilon^{\delta}_z = 2 \, t^{\delta}_{z} \cos(k_z)$ for the $d_{xy}$ orbital
with $t_z = -0.80$, and $t^{\delta}_z = 0.15$.

\section{Deriving J$_{\text{eff}}=1/2$ basis hoppings from t$_{2g}$ hoppings}
\label{app:t2gtoj}
Consider a single bond with t$_{2g}$ orbitals on each site with the overlaps between the various orbital combinations (these are diagonal in spin space) given by
$t_{yz}$, $t_{xz}$, and $t_{xy}$ for the intra-orbital hoppings and $t_{yz-xz}$, $t_{xz-yz}$, $t_{yz-xy}$, $t_{xy-yz}$, $t_{xz-xy}$, and $t_{xy-xz}$ for the inter-orbital hoppings. The J$_{\text{eff}}=1/2$ orbitals are given by
$\ket{J_z=\pm\frac{1}{2}} = \frac{1}{\sqrt{3}}(\ket{d_{yz},\mp s} \pm i\ket{d_{xz},\mp s} \pm \ket{d_{xy},\pm s}).$
and the time-reversal invariant hopping Hamiltonian for a bond is of the form $t \sigma^0 + i \sum_{\alpha = x, y, z} t^{\alpha} \sigma^{\alpha}$. From the $t_{2g}$ overlaps, the parameters can be obtained as
\begin{align}
\label{Eq:j12hopping}
t &= \frac{1}{3} (t_{yz} + t_{xz} + t_{xy}) \\
 t^z &= \frac{1}{3} (t_{yz-xz} - t_{xz-yz}) \\
t^y &= \frac{1}{3} (t_{xy-yz} - t_{yz-xy}) \\
 t^x &= \frac{1}{3} (t_{xz-xy} - t_{xy-xz})  
\end{align}

\subsection{Dispersions in J$_{\text{eff}}=1/2$ basis}
{\it Sr-214} -
\label{app:jhalf-214}
For Sr-214, in Sec. \ref{sec:Model}, we defined a Hamiltonian in Eq. \ref{eq:SLHamiltonian} matrices $A_k$, $B_k$, and $C_k$. 
$A_k$ describing hoppings within a layer was written as $\epsilon_k^{a} + \epsilon_k^{ad} \tau^x + \epsilon'^{ad}_k \tau^y\sigma^z$ with $\tau$ and $\sigma$ representing the sublattice and pseudo-spin degrees of freedom respectively. The dispersions in $A_k$ are
\begin{align*}
  \epsilon_k^{a}		&= 4t_n\cos(k_x)\cos(k_y) + 2t_{nn}(\cos(2k_x)+cos(2k_y)) \\
  \epsilon_k^{ad} 	&= 2t_p(\cos(k_x)+\cos(k_y)) \\
  \epsilon^{\prime ad}_k 	&= 2t^{\prime}_p(\cos(k_x)+\cos(k_y)).
\end{align*}
In the above, $\epsilon_k^{a}$, and $\epsilon_k^{ad}$ are pseudo-spin preserving hopping between the same sublattice (next-nearest and next-next-nearest) and different sublattices respectively. $t_n$ and $t_{nn}$ are next-nearest and next-next-nearest neighbour overlaps orbitals and $t_p$ is the nearest-neighbour overlap. 
$\epsilon_k^{\prime ad}$ is the pseudo-spin $\sigma^z$ dependent hopping between different sublattices with overlap $t^{\prime}_p$. The parameter set used was $(t_p,t'_p,t_n, t_{nn}) = (-0.57, 0.10,-0.067,0.033)$.

The inter-layer hoppings between layers 1 and 2 (also 3 and 4) were denoted by $B_k = \epsilon_k^{b} + \epsilon_k^{bd} \tau_x + 
			\epsilon_k^{bz} \tau_y\sigma_z + \epsilon_k^{by} \tau_y\sigma_y + \epsilon_k^{bx} \tau_y\sigma_x$ with the dispersions given by
\begin{align}
  \epsilon_k^{b} 		&= 2t_{i}\cos((k_x + k_y)/2) \\
  \epsilon_k^{bd} 	&= 2t_{id}\cos((k_x - k_y)/2) \\
  \epsilon_k^{bz} 	&= 2t_{i}^{z}\cos((k_x - k_y)/2) \\
  \epsilon_k^{by} 	&= i 2t_{i}^{y}\sin((k_x - k_y)/2) \\
  \epsilon_k^{bx} 	&= i 2t_{i}^{x}\sin((k_x - k_y)/2).
\end{align}
Here, $t_i$ is the hopping between the same sublattice and same $J_z$ while $t_{id}$ is the hopping between different sublattice with same $J_z$. The superscript $x$ in $t_{i}^{x}$, for example,  represents a pseudospin $\sigma^{x}$ dependent hopping between two orbitals on different sublattices. 
The remaining inter-layer hoppings between the second and third layers (also fourth and first layers) were captured by $C_k = \epsilon_k^{c} + \epsilon_k^{cd} \tau_x + 
\epsilon_k^{cz} \tau_y\sigma_z + \epsilon_k^{cy} \tau_y\sigma_y + \epsilon_k^{cx} \tau_y\sigma_x$ with the dispersions given by
\begin{align}
  \epsilon_k^{c} 		&= 2t_{i}\cos((k_x - k_y)/2) \\
  \epsilon_k^{cd} 	&= 2t_{id}\cos((k_x + k_y)/2) \\
  \epsilon_k^{cz} 	&= 2t_{i}^{z}\cos((k_x + k_y)/2) \\
  \epsilon_k^{cy} 	&= -i 2t_{i}^{y}\sin((k_x + k_y)/2) \\
  \epsilon_k^{cx} 	&= i 2t_{i}^{x}\sin((k_x + k_y)/2).
\end{align}
These inter-layer overlap parameters were obtained by a fit to the splitting of the bands seen in LDA calculations of Ref. \onlinecite{Martins:2011fk}. The parameters used were $(t_i,t_{id},t_{iz},t_{iy},t_{ix}) = (-0.029,-0.0275,-0.0135,-0.0095,0.0095)$.

{\it Sr-327} -
\label{app:jhalf-327}
For Sr-327, the matrices $A_k$ and $C_k$ are the same as those for Sr-214, but $B_k$ differs because of the bilayer nature.
$B_k$ has the form  $t_c \tau_x + t^{\prime}_c\tau_y\sigma_x$ as described in Sec. \ref{sub:sr__3_ir__2_o__7_model}.
These bilayer hoppings are given by $(t_c,t^{\prime}_c) = (-0.48,0.24)$ in units of $t$. These parameters can be obtained from overlaps in the t$_{2g}$ basis described in App. \ref{app:t2g-327} as $t_c = \frac{1}{3}(2t_z + t^{\delta}_{z})$ and $t^{\prime}_c = \frac{2}{3}t^{\prime}_z$ .


\bibliography{BibSrIrO}

\end{document}